%% file: main.tex
\documentclass[reprint,nofootinbib,
aps,superscriptaddress,twocolumn,showpacs,longbibliography]{revtex4-2}
\usepackage{graphicx,color}
\usepackage{amsmath,amsfonts,enumerate,amsthm,amssymb,bbm}
\usepackage[colorlinks=true,citecolor=blue,linkcolor=magenta]{hyperref}
\usepackage{multirow}
\usepackage{bbold}
\usepackage{braket}
\usepackage{soul}
\usepackage[caption = false]{subfig}
\usepackage{xcolor}
\usepackage{dsfont}
\usepackage{nicefrac}
\usepackage[linesnumbered,ruled,vlined]{algorithm2e}
\usepackage{quantikz}
\usepackage{apptools}

\mathchardef\ordinarycolon\mathcode`\:
\mathcode`\:=\string"8000
\begingroup \catcode`\:=\active
  \gdef:{\mathrel{\mathop\ordinarycolon}}
\endgroup

\theoremstyle{plain}

\theoremstyle{definition}

\theoremstyle{remark}

\AtAppendix{\counterwithin{thm}{section}}
\AtAppendix{\counterwithin{corol}{section}}
\AtAppendix{\counterwithin{lem}{section}}
\AtAppendix{\counterwithin{propos}{section}}
\AtAppendix{\counterwithin{defn}{section}}
\AtAppendix{\counterwithin{rmk}{section}}
\AtAppendix{\counterwithin{ex}{section}}

\usepackage[toc,page]{appendix}

\begin{document}

\title{ Low $T$-count preparation of nuclear eigenstates with tensor networks }

\author{Joe Gibbs}
\affiliation{School of Mathematics and Physics, University of Surrey, Guildford, GU2 7XH, UK}
\affiliation{AWE, Aldermaston, Reading, RG7 4PR, UK}

\author{Lukasz Cincio}
\affiliation{Theoretical Division, Los Alamos National Laboratory, Los Alamos, NM 87545, USA}

\author{Chandan Sarma}
\affiliation{School of Mathematics and Physics, University of Surrey, Guildford, GU2 7XH, UK}

\author{Zo\"e Holmes}
\affiliation{Institute of Physics, Ecole Polytechnique Fédéderale de Lausanne (EPFL), CH-1015, Lausanne, Switzerland}
\affiliation{Centre for Quantum Science and Engineering, École Polytechnique Fédérale de Lausanne (EPFL), Lausanne, Switzerland}

\author{Paul Stevenson}
\affiliation{School of Mathematics and Physics, University of Surrey, Guildford, GU2 7XH, UK}
\affiliation{AWE, Aldermaston, Reading, RG7 4PR, UK}

\date{\today}

\begin{abstract}

We present an efficient protocol leveraging classical computation to support Initial State Preparation for strongly correlated fermionic systems, a critical bottleneck for fault-tolerant quantum simulation. Focusing on nuclear shell model eigenstates, we first demonstrate that the Density Matrix Renormalization Group algorithm can efficiently approximate target states as Matrix Product States, capitalizing on the favourable entanglement structure of these fermionic systems. These high-fidelity approximations are then leveraged as a classical resource in a variational circuit optimization scheme to compile shallow quantum circuits. We establish concrete resource estimates by decomposing the resulting circuits into the industry-standard Clifford$+T$ gateset, exploring the benefits of specialized $U3$ synthesis techniques. 
For all nuclear systems tested, on up to 76 qubit Hamiltonians, we consistently find low $T$-count circuits preparing the nuclear eigenstates to high fidelity with $\sim 2\times 10^4$ total $T$ gates.
This low number gives confidence these eigenstates can be prepared on early fault-tolerant quantum computers. 
Our work establishes a viable path toward practical ground state preparation for nuclear structure and other fermionic applications.

\end{abstract}

\maketitle

\section{Introduction}

Simulating strongly correlated fermionic systems, such as atomic nuclei, is a grand challenge in modern computational physics. 
The complexity arises from the exponential growth of the Hilbert space with particle number, rapidly rendering exact diagonalization intractable. The quantum simulation of nuclear physics is central to understanding the origins and dynamics of matter itself. 
Accurately simulating these quantum many-body systems from first principles, a task that remains largely beyond the capability of classical computers, would provide unprecedented insights to both fundamental physics and industrial applications.

Quantum Phase Estimation (QPE)~\cite{kitaev1995quantum} is a foundational quantum algorithm that holds the promise of determining the energy eigenvalues of these complex Hamiltonians with Heisenberg-limited precision, offering a critical path toward quantum advantage in nuclear physics, quantum chemistry, and materials science. The success probability of QPE is fundamentally dependent on the quality of the initial state by its squared overlap with the target eigenstate~\cite{poulin2018quantum}. This overhead can be improved to scale with the overlap magnitude by amplitude amplification techniques~\cite{ge2019faster}.
While early studies assumed access to an exact eigenstate, preparing a sufficiently accurate approximate eigenstate 
is now recognized as a significant complexity bottleneck of fault-tolerant QPE~\cite{lee2023evaluating}. 
Due to the overheads associated with magic-state distillation, it is expected that the number of non-Clifford gates will be a dominant source of error in fault-tolerant quantum computations.
Strategies for Initial State Preparation include, adiabatic state preparation~\cite{costa2025quantum,costa2025quasiparticle}, preparing a sum of slater determinants~\cite{fomichev2024initial}, and the Variational Quantum Eigensolver (VQE)~\cite{carrasco2025comparison, yoshida2025bridging}; each present significant resource trade-offs, underscoring the urgent need for resource-efficient compilation techniques.

The challenge of designing quantum resource-efficient circuits necessitates hybrid quantum-classical approaches. Here, powerful classical computation is leveraged to inform the design of low-depth, high-fidelity quantum circuits. Among these approaches, Tensor Network methods~\cite{orus2014practical}, a variational ansatz for describing quantum states as a low-rank network of tensors, offer a promising direction. 
Despite the inherent one-dimensional structure of Matrix Product States (MPS)~\cite{schollwock2011density}, the power and stability of the Density Matrix Renormalization Group algorithm (DMRG) remains a leading method for studying two-dimensional quantum lattice systems~\cite{stoudenmire2012studying} and quantum chemistry~\cite{chan2008introduction}.
This classical tractability, combined with the acceleration by leveraging massive parallel processing~\cite{hyatt2019dmrg, ganahl2023density, menczer2024two, brower2025mixed}, positions Tensor Networks as a potent tool for both classical simulation and quantum algorithm design.

\begin{figure*}[t!]
    \centering
    \includegraphics[width=\linewidth]{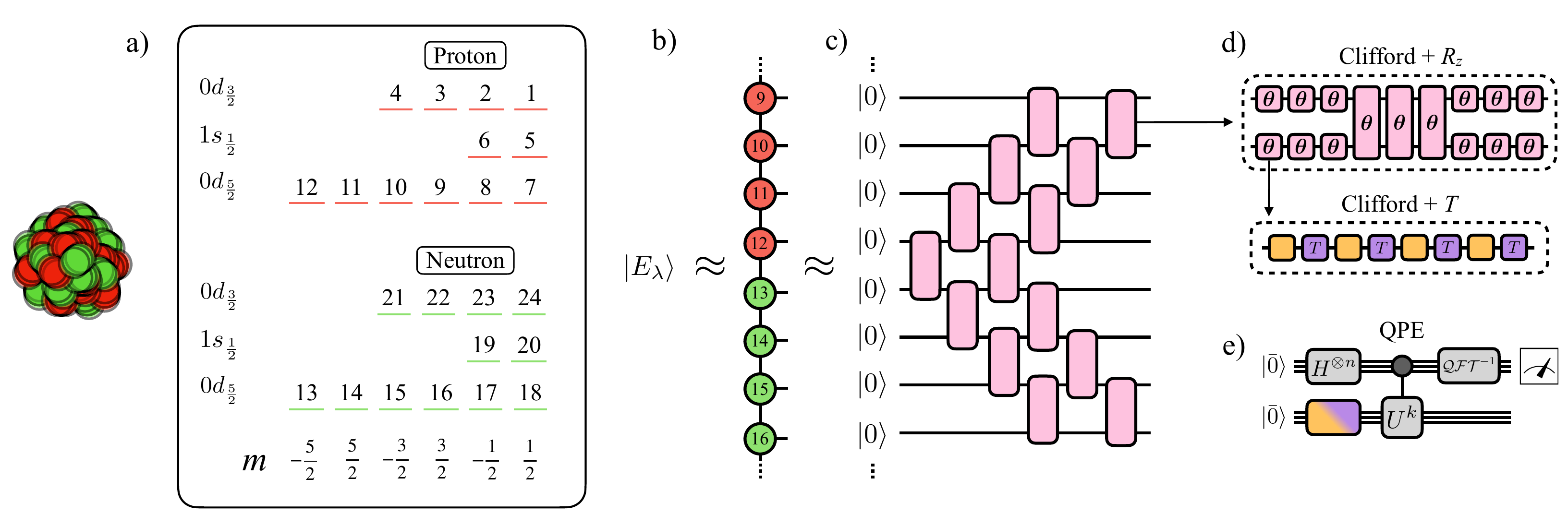}
    \caption{\textbf{Overview.} Sketch of method to approximately prepare nuclear shell model eigenstates on fault-tolerant quantum computers, aided by tensor networks. 
    a) Nuclear shell model Hamiltonians are defined on a valence space of proton and neutron orbitals. Here this is visualized for a 24 qubit example. 
    b) We approximately represent a target nuclear eigenstate, $|E_\lambda\rangle$, as an MPS using DMRG, exploiting known entanglement structures in nuclei. 
    c) These MPS are used as a resource in a variational circuit optimization, maximizing fidelity with the target MPS. 
    d) These circuits have a simple representation in Clifford$+R_z$ gates. A further decomposition by unitary synthesis finally gives circuits described in the Clifford$+T$ gateset.
    Across the nuclei studied, on up to 76 qubit Hamiltonians, we consistently find circuits preparing eigenstates to high fidelities with low $T$-counts of $\sim2\times 10^4$ total $T$ gates.
    e) These circuits are ready to prepare the initial state in future fault-tolerant quantum algorithms.}
    \label{fig:Overview}
\end{figure*}

In recent years, utilizing Tensor Networks for quantum circuit synthesis has become an active area of research. An established strategy is analytic decomposition, based on the ``sequential generation" scheme pioneered in Ref.~\cite{schon2005sequential}. This method \textit{deterministically} compiles the MPS by viewing it as the output of a sequential process. An ancillary system, whose dimension $D$ is equal to the MPS bond dimension $\chi$, is unitarily coupled to and entangled with each qubit iteratively. 
While this scheme is exact in principle, its fault-tolerant resource cost has been studied in ~\cite{fomichev2024initial}, and still presents a critical bottleneck.
This approach was combined with a method for synthesising unitaries with low Toffoli count by decomposing the unitary into a sequence of diagonal phasing operations in~\cite{berry2025rapid}. For a challenging, strongly correlated system like the FeMo cofactor, it was estimated to require up to $1.36 \times 10^{9}$ Toffoli gates to prepare a high-fidelity initial state on 833 qubits.
Thus deterministic methods seemingly lead to very deep circuits, highlighting the need for new approaches to compiling circuits to prepare initial states. 

The alternate strategy to the analytic decomposition is an \textit{approximate variational} compilation. Instead of a direct, costly translation, this hybrid approach leverages the MPS as an efficient classical target. A low-resource quantum circuit ansatz is variationally optimized using classical computation to maximize its fidelity with the target MPS, prioritizing quantum resource efficiency. 
The direct and exact translation is traded for variational freedom and flexibility in choosing a circuit ansatz design. There is a growing body of work investigating methods for performing this variational compilation from MPS to quantum circuit~\cite{lin2021real, ran2020encoding, rudolph2022decomposition, gibbs2024deep, chai2025resource, gibbs2025learning, jaderberg2025variational, jamet2023anderson, anselme2024combining, rudolph2022synergy, le2025riemannian, iaconis2024quantum, kanno2025tensor, szoldra2026scalable, ballarin2025efficient}. Typically these works have focused on supporting quantum algorithms running on near-term devices, where minimising 2-qubit gate counts is the primary concern as the dominant source of errors. However, when targetting applications on fault-tolerant quantum computers, these methods should be specialized to reduce non-Clifford gate counts.

In this paper we develop methods of MPS-based compilation of circuits with low non-Clifford gate counts to the simulation of nuclear structure, a domain particularly suited for quantum computation due to the clean mapping between nuclear orbitals and qubits. 
Furthermore, we exploit the specific entanglement structure of atomic nuclei, characterized by weak proton-neutron correlations compared to strong pairing interactions, to construct highly efficient Tensor Network representations.
While the simulation of nuclear structure is a key target for quantum advantage, the use of Tensor Network methods like DMRG to approximate nuclear shell model eigenstates classically remains nascent~\cite{dukelsky2001new, pittel2006density, legeza2015advanced, fossez2022density, tichai2023combining}. 

\medskip

Our work proceeds as follows (and is summarized in Fig.~\ref{fig:Overview}). In Sec.~\ref{sec:MPS_pre_proc} we assess the suitability of the DMRG algorithm in computing MPS approximations of nuclear shell model eigenstates. We show that overlap magnitudes with the exact ground states increases rapidly with increasing bond dimension.
This establishes a `crossover' regime where classical methods are sufficient to guide the quantum computer, but the quantum processor is required for the final Heisenberg-limited phase estimation.
In Sec.~\ref{sec:compilation}, we describe the methods for variational circuit compilation to transform these high-fidelity MPS targets into shallow quantum circuits, optimized for an efficient representation by the Clifford$+R_z$ gateset. We then give specialized techniques, leveraging new MPS-based techniques for unitary synthesis~\cite{hao2025reducing}, to reduce overheads for the decomposition to the Clifford$+T$ gateset.
Finally, in Sec~\ref{sec:final_test}, we apply these techniques to systems beyond exact simulation, to 76 qubit Nuclear Hamiltonians. To prepare these eigenstates with high overlap magnitudes, we find only $\sim 2\times 10^4$ $T$ gates are required, giving confidence these can be prepared on early fault-tolerant quantum computers.

These low resource estimates for approximately preparing nuclear eigenstates, through the novel dual use of tensor networks for both the circuit optimization and unitary synthesis, provides a new and promising approach to initial state preparation of strongly correlated fermionic systems on fault-tolerant quantum computers.

\newpage

\section{MPS Pre-Processing} \label{sec:MPS_pre_proc}

\subsection{Hamiltonian}\label{sec:hamiltonian}

We first describe the nuclear systems our techniques are benchmarked against. The nuclear shell-model m-scheme \cite{deshalittalmi} Hamiltonian expressed in second quantized form is given as
\begin{equation}
    H = \sum_i \epsilon_i a^\dagger_ia_i + \frac{1}{2}\sum_{i,j,k,l} V_{ijkl} a^\dagger_i a^\dagger_j a_k a_l, 
\end{equation}
where the operators $a^\dagger_i$ and $a_i$ respectively correspond to creation and annihilation operators of a nucleon in the $i^\text{th}$ single-particle orbital; in a quantum simulation each nucleon single-particle orbital is mapped onto one qubit. 

\medskip

The coefficients $\epsilon_i$ and $V_{i,j,k,l}$ correspond to single-particle energies and two body matrix elements respectively, where the basis is explicitly diagonal in the one-body part.  Each single-particle state is characterized by quantum numbers $\{t_z,n,l,j,j_z\}$, where $t_z$ is the isospin projection, $n$ and $l$ represent the radial and orbital angular momentum quantum numbers, $j$ and $j_z$ give the total angular momentum and its projection along the z-axis respectively~\cite{sarma_low-circuit-depth_2026}.

We first perform extensive testing for a range of nuclear eigenstates within the limits of exact classical representation. We work with a nuclear shell model Hamiltonian, with the \textit{usdb} interaction \cite{PhysRevC.74.034315}. The model space consists of the $0d_\frac{5}{2}, 1s_\frac{1}{2}, 0d_\frac{3}{2}$ orbitals for both protons and neutrons above an inert $^{16}$O core, with 24 single particle orbitals in total  (corresponding to 24 qubits).  This is the model space mapped out in Figure \ref{fig:Overview}a).
We target different numbers of valence protons and neutrons in the range $(n_p,n_n)\in\{(2,2),(2,3),(2,4),(3,3),(3,4),(3,5)\}$, corresponding to the nuclei 
$\{^{20}$Ne, $^{21}$Ne, $^{22}$Ne, $^{22}${Na}, $^{23}${Na}, $^{24}${Na}$\}$ respectively. 
The Hamiltonians describing these nuclear isotopes are easily within the limits of exact diagonalisation, allowing exact verification of errors on eigenenergies and exact calculation of fidelities. 

To go beyond the limits of exact simulation, in Sec.~\ref{sec:final_test} we study the nuclei $^{142}$Ce and $^{143}$Ce using the KHHE \cite{PhysRevC.44.233} shell model interaction. This effective interaction was constructed for the model space 50 $\leq$ $Z$ $\leq$ 82 and 82 $\leq$ N $\leq$ 126 above the inert $^{132}_{50}$Sn core. It involves 11 orbitals: five proton orbitals (2$s_{1/2}$, 0$g_{7/2}$, 0$h_{11/2}$, 1$d_{5/2}$, 1$d_{3/2}$ ) and six neutron orbitals (0$h_{9/2}$, 0$i_{13/2}$, 1$f_{7/2}$, 2$p_{3/2}$, 1$f_{5/2}$, 2$p_{1/2}$ ). In total there are 76 single-particle orbitals, mapping onto the same number of qubits. Within this valence space $^{142}$Ce ($^{143}$Ce) is defined as a 10 (11) valence nucleon system having eight valence protons and two (three) valence neutrons.

\newpage
\subsection{DMRG }

Here we describe and study the application of the 2-site DMRG algorithm to prepare eigenstates of these Hamiltonians as MPS, using the $\texttt{ITensorMPS}$ package~\cite{fishman2022itensor}.

To circumvent the challenges from exponential growth of the Hilbert space with system size, one seeks compact representations efficiently capturing physically relevant states. Matrix Product States~\cite{schollwock2011density} provide such a representation by exploiting the classical simulability of weakly entangled states~\cite{Vidal2003Efficient}. A quantum state of $L$ sites is expressed as a product of local tensors connected by virtual bonds. Explicitly this can be written as 
$$
|\psi\rangle = \sum_{s_1, s_2, \dots, s_L} A_1^{s_1} A_2^{s_2} \cdots A_L^{s_L} |s_1 s_2 \dots s_L\rangle,
$$
where each $A_i^{s_i}$ is a complex matrix of size $\chi_{i-1}\times \chi_i$, and $s_i$ labels the local basis states of site $i$. The integers $\chi_i$ are referred to as bond dimensions, controlling the expressiveness of the MPS and determining the maximum amount of bipartite entanglement the state can represent. Open boundary conditions are fixed with $\chi_0=\chi_L=1$.
The DMRG algorithm~\cite{white1992density} is a tensor network eigensolver, efficiently updating the tensors in a variational MPS representing the state $|\psi\rangle$ to minimize the variational energy $E(|\psi\rangle) = \langle\psi|H|\psi\rangle / \langle\psi|\psi\rangle$ with respect to the target Hamiltonian $H$~\cite{schollwock2011density}. The 2-site variant of DMRG allows adaptive truncations on tensor decompositions to allow the bond dimension to flexibly vary in size to accommodate a target accuracy.

\medskip

Each single-particle orbital in the target nuclear system can be mapped onto a single qubit, and these qubits are individually represented by tensors on the MPS. We have a freedom in choosing the mapping from orbitals to MPS tensors.
A poorly chosen mapping not matching the underlying correlation structure may cause unnecessary increases in bond dimensions to represent the target state. 
For example, if two single-particle orbitals are strongly correlated and their mapped qubits are on separate ends of the MPS chain, this will require long-range entanglement to be modelled and increase the bond dimension required to represent the quantum state accurately.
Previous work~\cite{tichai2023combining, perez2023quantum, brokemeier2025quantum, johnson2023proton} has studied correlations between these orbitals in nuclear shell model ground states. Notable features include significantly weaker proton-neutron correlations compared to proton-proton or neutron-neutron correlations. It was also observed that there were strong correlations between single particle orbitals with opposite $j_z$ quantum numbers, as expected from the strong pairing residual interaction in nuclei, acting between time-reversed states \cite{RevModPhys.75.607}.  

We take advantage of the above observations in choosing our mapping; the orbitals are ordered in increasing single particle energies, and within the orbitals the single particle orbitals are ordered in decreasing $|j_z|$ values, so that single particle orbitals with opposite $j_z$ quantum numbers are always represented by neighbouring MPS tensors. The observation of weak correlation between proton and neutron orbitals suggests separating the proton and neutron orbitals onto separate halves of the MPS. 

\medskip

The eigenstates we target have a specified number of protons and neutrons within the valence space. This corresponds to finding the eigenstate within a $U(1)_\text{proton}\times U(1)_\text{neutron}$ symmetry sector of the full Hamiltonian. We therefore use $U(1)\times U(1)$ symmetry-preserving MPS, which are able to take advantage of the reduced memory required to express states in this restricted Hilbert space.

\begin{figure}[t]
    \centering
    \includegraphics[width=\columnwidth]{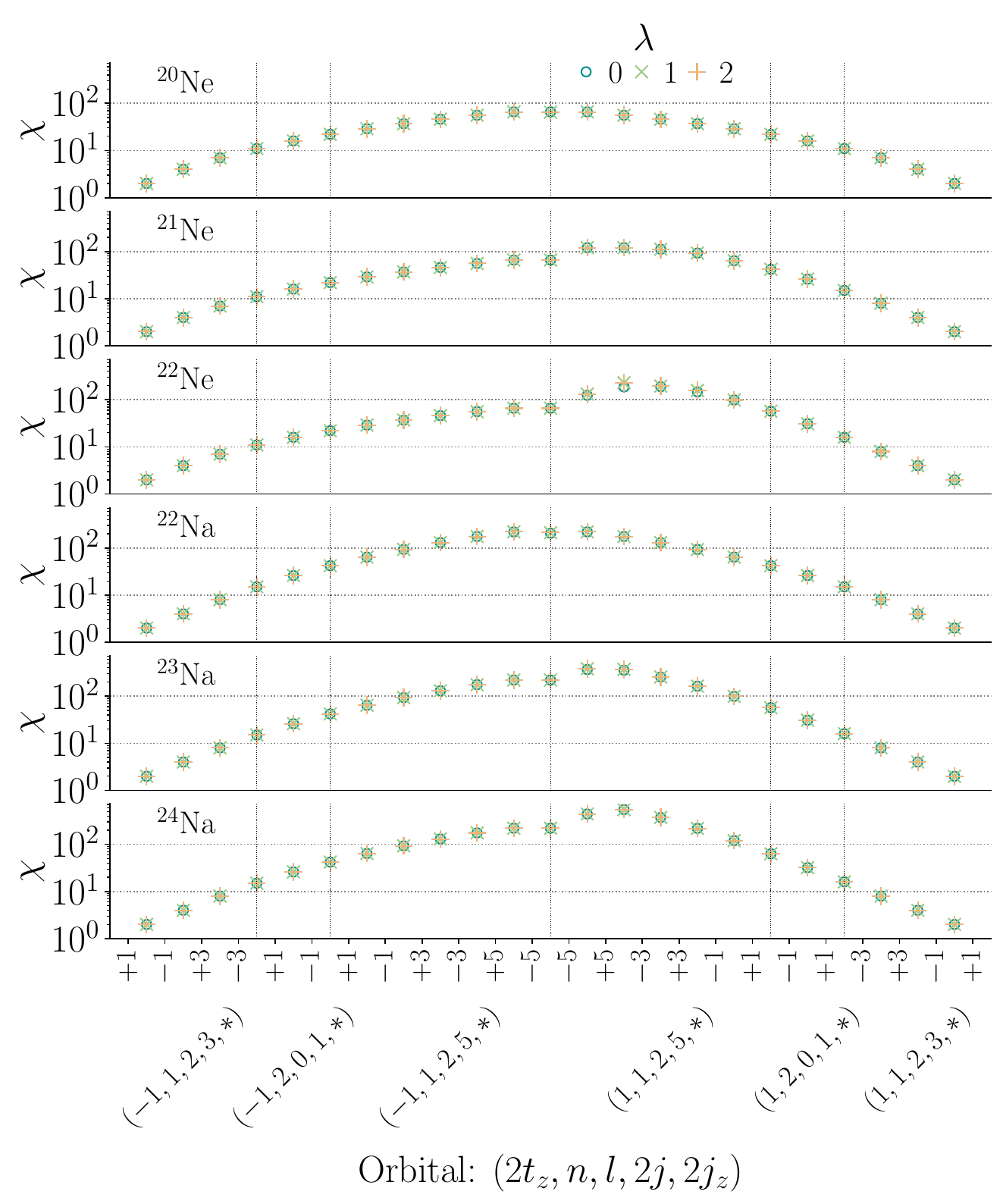}
    \caption{\textbf{Neon and Sodium - MPS Bond Dimension Distribution.} For the three neon isotopes $^{(20-22)}$Ne and three sodium isotopes $^{(22-24)}$Na, we apply perform a high-accuracy DMRG calculation to determine the three lowest-lying eigenstates (enumerated by $\lambda\in\{0,1,2\}$). The x-axis indicates the MPS site tensors, labelled by the quantum numbers of single-particle orbital they represent. The y-axis measures the virtual bond dimension between these site tensors in the outputted MPS, after DMRG is performed with a minimum singular value of $10^{-8}$ retained after tensor decompositions.}    
    \label{fig:usdb_dmrg_chis}
\end{figure}

We apply the 2-site DMRG to these nuclear Hamiltonians, with the initial state set to a random MPS within the target symmetry sector.
Our methods are not restricted to studying ground states. Therefore as well as the ground state, we target the first two excited states, and we index these first three eigenstates by $\lambda \in \{0,1,2\}$.
We produce excited states using DMRG by modifying the Hamiltonian to penalise overlap with previously found approximate eigenstates $|\psi_\lambda\rangle$, by $H_\lambda = H + w\sum_{i=0}^{\lambda-1} |\psi_i\rangle\langle\psi_i|$~\cite{schollwock2011density}. This method requires the weight of this penalty term, $w$, to be larger than the energy gap between eigenstates; we set this to 20 MeV throughout our results.

For all DMRG results presented in this section, we verify their correctness by comparing the variational energy values to exact diagonalisation calculations using the $\texttt{QuSpin}$ library~\cite{weinberg2017quspin}, similarly exploiting the $U(1)_\text{proton}\times U(1)_\text{neutron}$ symmetry structure. The energy values all agreed to within a relative error of $10^{-5}$.

We now study the ability of MPS to represent the first three eigenstates of the $^{20-22}$Na and $^{22-24}$Ne isotope test cases to high accuracy.
The target Hamiltonian, described in Sec.~\ref{sec:hamiltonian}, has 12 single particle orbitals, requiring 24 qubits to represent both proton and neutron orbitals. 
The results of the DMRG calculation are shown in Fig.~\ref{fig:usdb_dmrg_chis}.
We apply DMRG with an unrestricted maximum bond dimension, with singular value magnitudes less than $10^{-8}$ truncated during tensor decompositions.
There is a remarkable uniformity in the bond dimension of the MPS outputted by the DMRG across the three eigenstates for each nucleus. 
An increase in bond dimension required to express these eigenstates accurately is observed for increasing nucleon numbers, and is noticeable by the symmetric bond dimension distribution for $n_p = n_n$, with increasing bond dimension over the MPS tensors representing the neutron orbitals when $n_n > n_p$.

\subsection{Approximate MPS Representation}

In the previous section, we studied the ability of DMRG to prepare the low-lying eigenstates of a range of nuclear isotopes. We leave open the question of DMRG as an effective eigensolver for high accuracy simulation of nuclear eigenstates, which would require more extensive studies.
For our target application, the compilation of circuits to aid Initial State Preparation on fault-tolerant quantum computers, we have more relaxed constraints. Here the MPS produced by DMRG only need to capture high fidelity with the exact eigenstate, rather than computing eigenenergies to many decimal places.

This suggests the following question. For a target eigenstate beyond the limits of exact representation by MPS, how well can a low bond dimension MPS $|\Phi(\chi) \rangle$ approximate the exact eigenstate $|E_\lambda\rangle$, as measured by the overlap magnitude $\left| \langle \Phi(\chi) |E_\lambda\rangle \right|$. If tractable MPS can capture high fidelities, this would then be a valuable classical resource to input into a circuit compilation, to generate initial states for fault-tolerant quantum simulations. 

\begin{figure}[t]
    \centering    \includegraphics[width=\columnwidth]{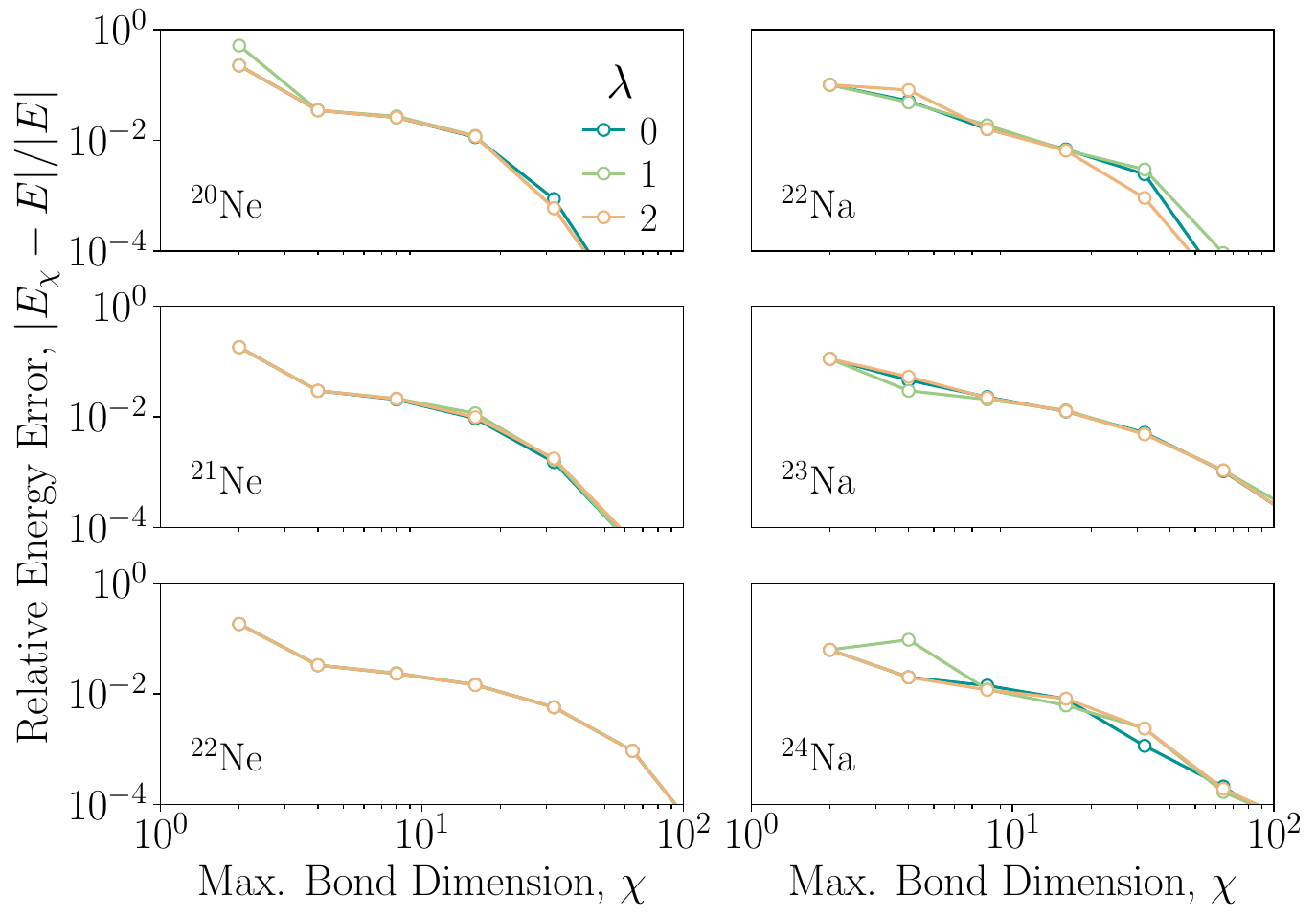}
    \caption{\textbf{Restricted bond dimension DMRG.} Convergence of variational energy computed by DMRG with respect to the exact value, as the maximum allowed bond dimension of the outputted MPS increases.}
    \label{fig:dmrg-smallchi}
\end{figure}

We repeat the DMRG procedure for computing the eigenstates as MPS, however now we restrict the maximum bond dimension and test how the variational energy values converge at this maximum bond dimension is increased. These results are shown in Fig.~\ref{fig:dmrg-smallchi}. For each eigenstate, compared to the reference energy value $E$ computed by exact diagonalisation, we compute the relative energy error $|E_\chi-E|/|E|$ where $E_\chi$ is the converged variational energy of the DMRG restricted to maximum bond dimension $\chi$. We find a fast reduction in this error with increasing bond dimension, with relative energy errors below $10^{-3}$ for all eigenstates found at $\chi < 100$.

Estimates of eigenenergies and energy gaps can be used to infer lower-bounds of overlaps of MPS with exact eigenstates~\cite{lin2025bounds}. 
Alternate extrapolation procedures for estimating the overlap have been proposed~\cite{berry2025rapid}, based on empirical relationships observed between overlaps of varying bond dimensions MPS approximations of the target MPS, which we employ in the later section Sec.~\ref{sec:final_test} for target ground states beyond the limits of exact representation with MPS.
Given in this section we work in a regime where the target eigenstates can be exactly expressed as MPS, we can directly probe the ability of lower bond dimension MPS to capture high fidelity with the exact eigenstate. 

\begin{figure}[t]
    \centering
    \includegraphics[width=\columnwidth]{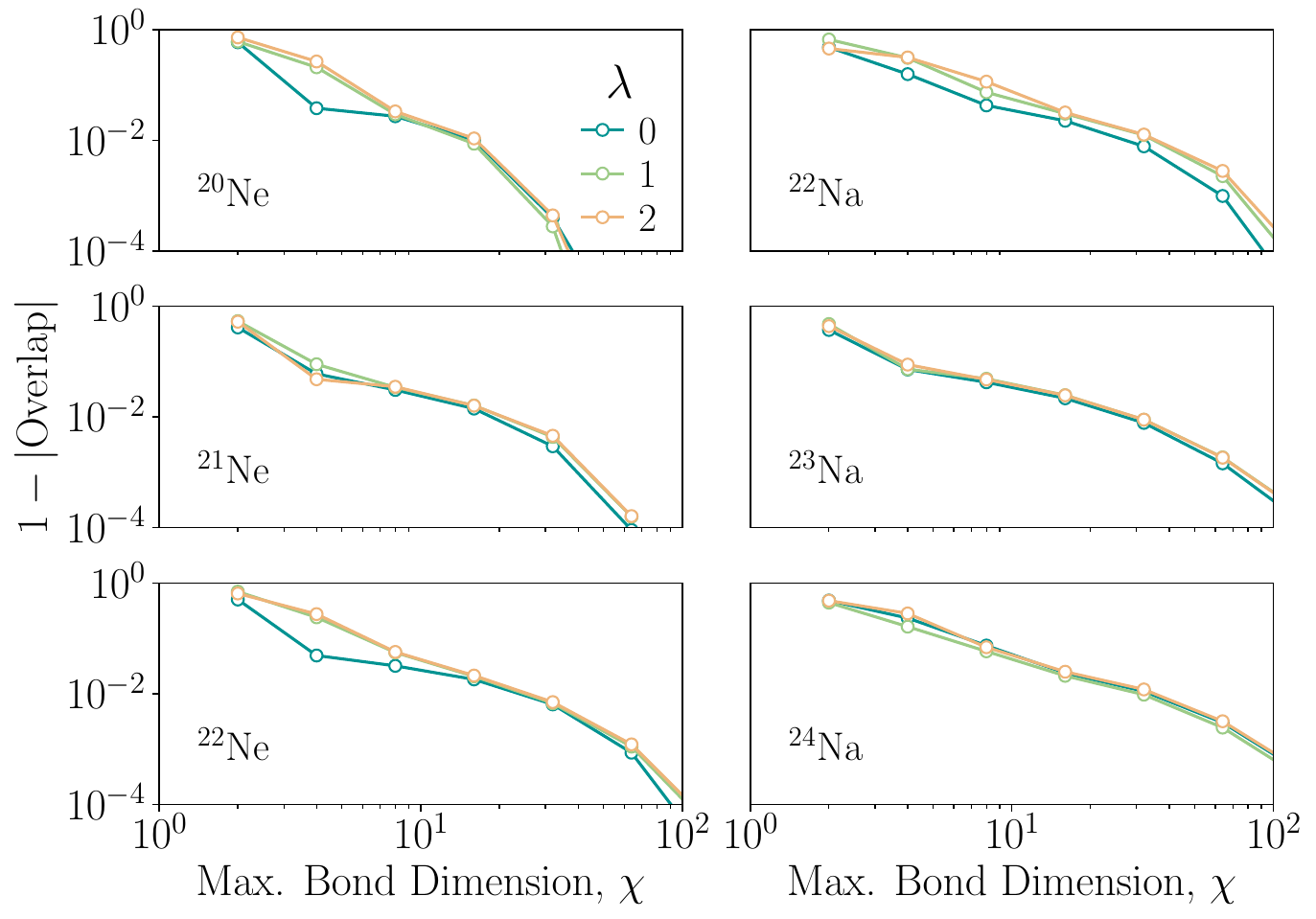}
    \caption{\textbf{MPS Compression.} Here we variationally compress the set of MPS shown in Fig.~\ref{fig:usdb_dmrg_chis} representing the states $|E_\lambda\rangle$ for different nuclei, producing the MPS $|\Phi(\chi)\rangle$ compressed to a maximum bond dimension of $\chi$. After the compression we compute the overlap error $1-\big|\langle\Phi(\chi)|E_\lambda\rangle\big|$, for increasing $\chi$, and find a rapid decrease in error to the exact eigenstate.}
    \label{fig:usdb_approx_mps}
\end{figure}

We study this by a variational compression~\cite{schollwock2011density} of the high accuracy MPS shown in Fig.~\ref{fig:usdb_dmrg_chis}, where the tensors of a variational MPS with a capped maximum bond dimension are swept back and forth with updates maximizing the overlap with the target MPS, we denote this converged MPS $|\Phi(\chi) \rangle$. 
We scan across increasing maximum bond dimensions $\chi$, finding the state $|\Phi(\chi) \rangle$ with maximum overlap with the true ground state. This data is shown in Fig.~\ref{fig:usdb_approx_mps}. We find these low dimension MPS to be highly effective in achieving high fidelity with the exact eigenstate. For all examples tested, a maximum bond dimension of 64 was sufficient for a high overlap magnitude of $0.99$.
This demonstrates that low bond dimension MPS, despite not necessarily being able to compute these variational energies with very high accuracy to the exact energy (e.g. to 8 decimal places), can still capture high fidelity with the exact eigenstate. 
This demonstrates that tensor network eigensolvers with restricted bond dimensions can yield MPS that serve as a valuable classical resource. These MPS can subsequently be used in an optimization process to learn quantum circuits for approximate eigenstate preparation.


\section{Circuit Compilation}\label{sec:compilation}

We now describe the compilation of circuits to approximately prepare these eigenstates. We intend for these circuits to serve as inputs for future fault-tolerant quantum computations, and our primary goal is to minimize the associated resource overhead. 

Fault-tolerant computations are typically expressed using a Clifford+$T$ gateset. Here, $T$ gates are expensive non-Clifford operations that must be minimized, as they require resource-intensive magic state distillation. Our compilation proceeds in two steps: first, we approximate the target state using circuits composed of Clifford+$R_z$ gates; second, we perform unitary synthesis to approximate the $R_z$ gates with Clifford+$T$ gates. We aim to achieve high overlap with the target state using circuits composed of as few $SU(4)$ unitaries as possible. This is because the required number of $R_z$ gates is proportional to the final $T$-count---the ultimate metric we wish to minimize. 

\subsection{MPS to Circuit Optimization}\label{sec:circuit_opt}

Here we describe the optimization of shallow circuits to approximately prepare the target eigenstates.
Our circuits are described by a sequence of 2-qubit $SU(4)$ unitaries acting between pairs of qubits. These unitaries are swept through and updated sequentially, by first computing their environment in the tensor network representing the overlap with the target MPS; a polar decomposition of this tensor gives the optimal $SU(4)$ unitary to update to maximise the overlap between the states. This is a standard practice in tensor network algorithms~\cite{evenbly2009algorithms}, and further used in MPS-based circuit optimizations~\cite{lin2021real, shirakawa2021automatic, rudolph2022decomposition, causer2023scalable, gibbs2024deep, chai2025resource}.
The gates are updated sequentially, where each gate in the circuit is updated in an order taking advantage of cheap updates to environment MPS when moving to the next gate. We sweep back and forth along the circuit, until the change in overlap magnitude has a relative error between sweeps of less than $10^{-4}$ .

\begin{figure}
    \centering
    \includegraphics[width=\columnwidth]{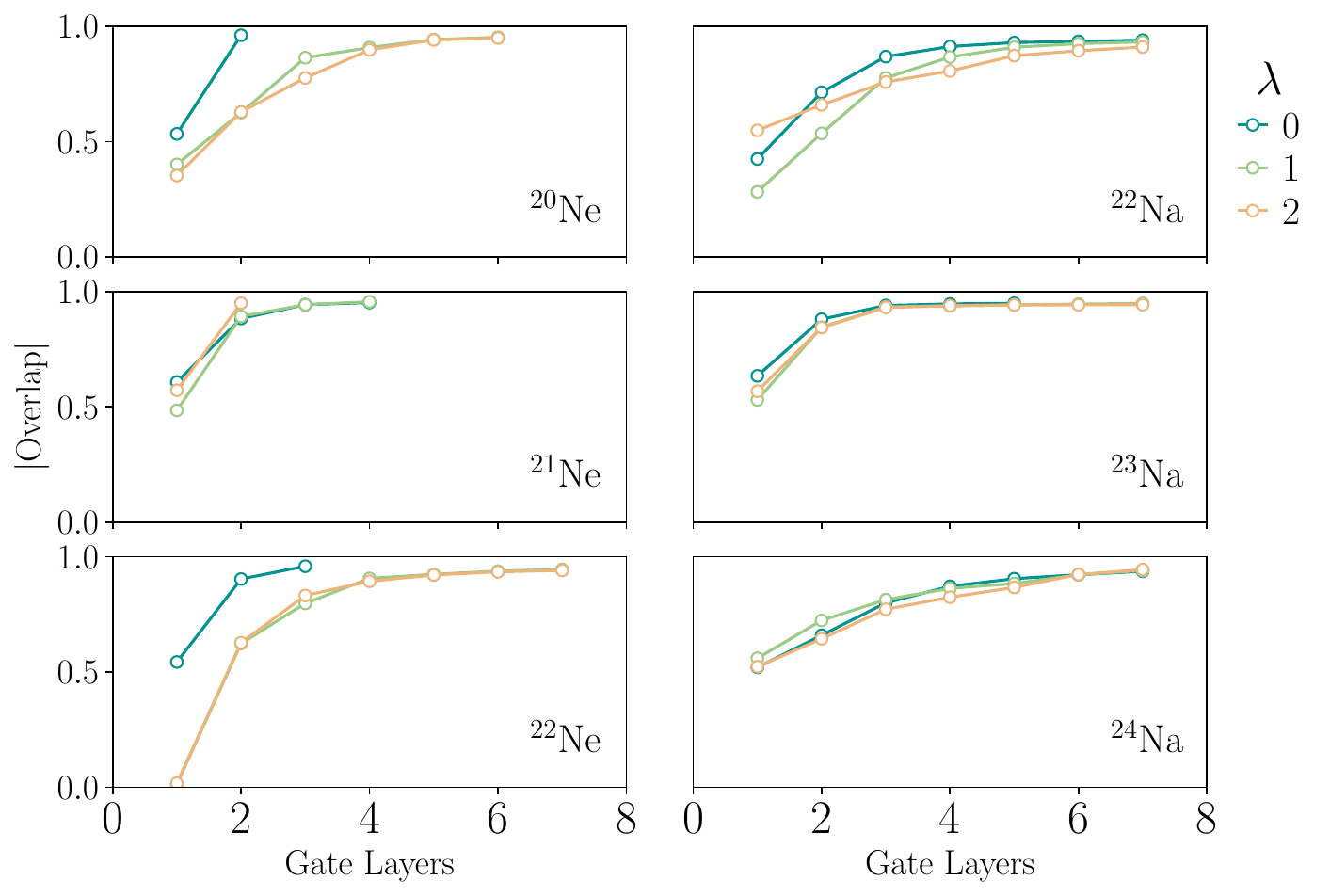}
    \caption{\textbf{Approximate state preparation with shallow circuits.} Short depth circuits, following the construction described in Sec.~\ref{sec:circuit_opt}, are optimized to maximize the overlap with the target eigenstate $|E_\lambda\rangle$. Here the overlap magnitude is computed by $\big|\langle E_\lambda|U|\mathbf{0}\rangle\big|$ where $U$ is the optimized circuit.}
    \label{fig:usdb_approx_circuit}
\end{figure}

Our circuits are composed of V-shaped staircase layers of $SU(4)$ unitaries, the most general 2-qubit unitary. 
Each layer starts with an $SU(4)$ unitary on the bond separating the proton and neutron orbitals, then a staircase of $SU(4)$ unitaries proceeds away from the centre on both sides. For our system with 24 qubits in total, this circuit of $L$ layers is specifically described by 
$$
\prod_{l=1}^L
\left(\prod_{i=11}^{1} U^l_{(i,i+1)}  \right)
\left(\prod_{i=12}^{23} U^l_{(i,i+1)} \right)
, 
$$
where $U^l_{(i,i+1)}$ are $SU(4)$ unitaries  acting between qubits $i$ and $i+1$.
A cartoon graphic of these layers is shown in Fig.~\ref{fig:Overview}c).

\begin{figure*}[t]
    \centering
    
    \includegraphics[width=0.9\linewidth]{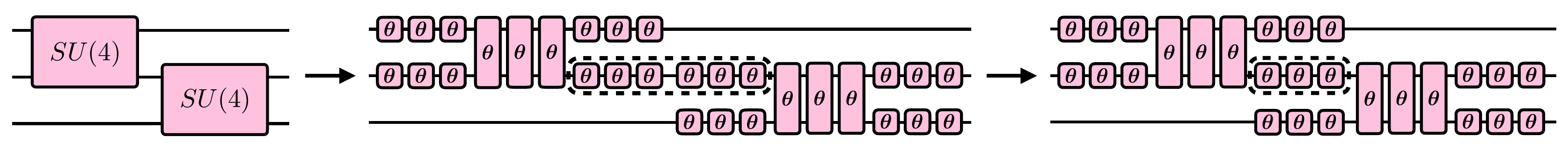}
    \caption{\textbf{Redundant Gate Removal.} After the variational circuit optimization, each $SU(4)$ unitary is converted into two qubit Pauli rotations sandwiched between arbitrary single qubit rotations, requiring at most 15 Pauli rotations. Due to the sequential application of these $SU(4)$ unitaries throughout the circuit, naively this results in sequences of 6 single qubit rotations acting consecutively, which can be trivially compressed down to at most 3 single qubit rotations, shown within the dashed lines. This occurs frequently throughout the circuit, on all qubits between gate layers, therefore significantly reducing the number of gates that require synthesizing to the Clifford$+T$ gateset. For all $SU(4)$ unitaries, other than the last layer, this reduces the number of Pauli rotations to synthesize from 15 to 9, a reduction of 40\%.}
    \label{fig:redundancy}
\end{figure*}

\begin{figure}[t]
    \centering
    
    \includegraphics[width=0.9\linewidth]{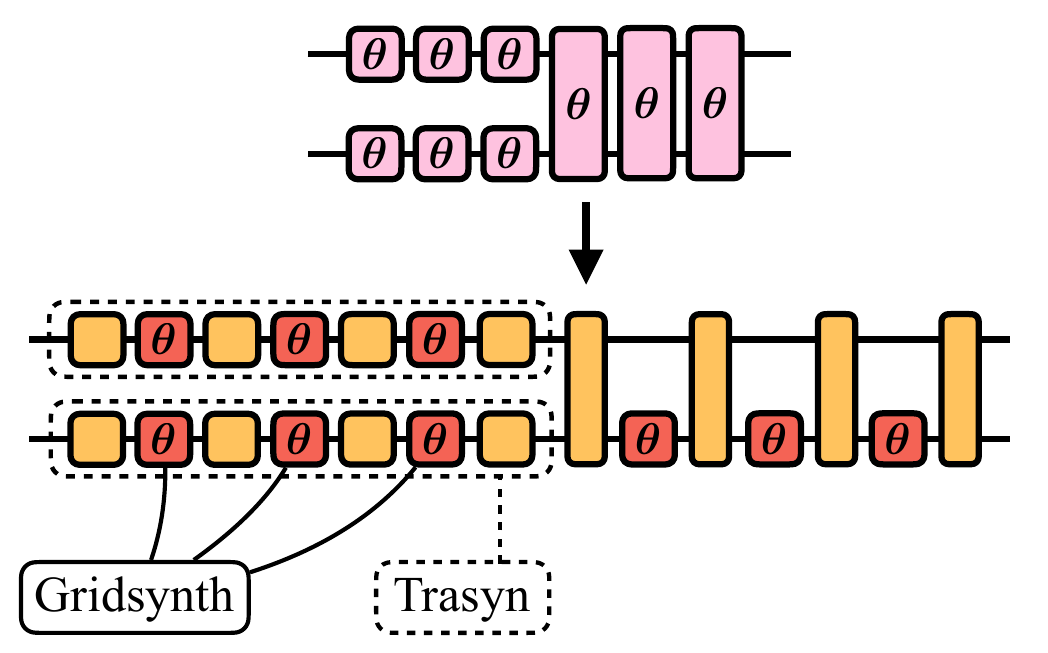}
    \caption{\textbf{Synthesis Methods.} After the circuit is expressed in terms of Pauli rotations (pink), these finally need synthesizing to the Clifford$+T$ gateset. First, we express these in terms of Clifford (orange) and $R_z$ (red) gates. The first synthesis approach uses the Gridsynth algorithm to decompose each $R_z$ gate into the Clifford$+T$ gateset. Frequently throughout the circuit, there are 3 consecutive $R_z$ gates (plus Clifford) on a single qubit, as shown within the dashed box, that are each decomposed by Gridsynth. This factor of 3 overhead can be removed if a more direct single qubit unitary synthesis routine is employed, and we use the Trasyn algorithm here. By exploiting this structure in our circuits, the hybrid strategy reduces the effective number of unitaries synthesized for this repeating gate block from 9 to 5, a reduction of 44.4\%.   }
    \label{fig:synthesis_methods}
\end{figure}

We use an optimization strategy that iteratively grows the circuit depth. A single layer ansatz is randomly initialised, and optimized until convergence. We then add a new layer initialised to maximise the decrease in infidelity. This is found by prepending a layer of identity gates, and with all gates in previous layers fixed, and we optimize over this new layer of gates until convergence. Then all gate layers in the circuit are optimized together until convergence again. This is similar to the $\text{Iter}[D_iO_\text{all}]$ protocol introduced in ~\cite{rudolph2022decomposition}; rather than truncating the residual state $V(\theta)^{\dagger}|\psi_\text{targ}\rangle$ down to a $\chi=2$ MPS, which permits an exact circuit preparation~\cite{ran2020encoding}, we instead optimize the new layer to maximise the overlap with the full residual state.


Fig.~\ref{fig:usdb_approx_circuit} shows the result of these optimizations, applied to the first three eigenstates of the nuclei $\{^{20}${Ne}, $^{21}${Ne}, $^{21}${Ne}, $^{22}${Na}, $^{23}${Na}, $^{24}${Na}$\}$. Each data point corresponds to a circuit $U$ optimized to approximately prepare the target eigenstate $|E_\lambda\rangle$, and for each circuit we plot the corresponding overlap $\big|\langle E_\lambda|U|\mathbf{0}\rangle\big|$. We find a rapid increase in fidelity as a function of gate layers, with around 5 gate layers sufficient for all nuclear eigenstates to have overlap magnitudes over 0.8.

Our goal is to generate circuits approximately preparing the target eigenstate with few $R_z$ gates. This is because a following unitary synthesis will result in a total $T$-count (the final metric we wish to minimize) proportional to the total number of $R_z$ gates. The circuit ansatz structure we used was somewhat generic, and could prepare any state with sufficient circuit depth; it is conceivable however, that a problem-inspired ansatz focusing on a physically relevant corner of the Hilbert space, may result in lower total $R_z$ gates.


\begin{figure*}[t]
    \centering
    \includegraphics[width=0.9\linewidth]{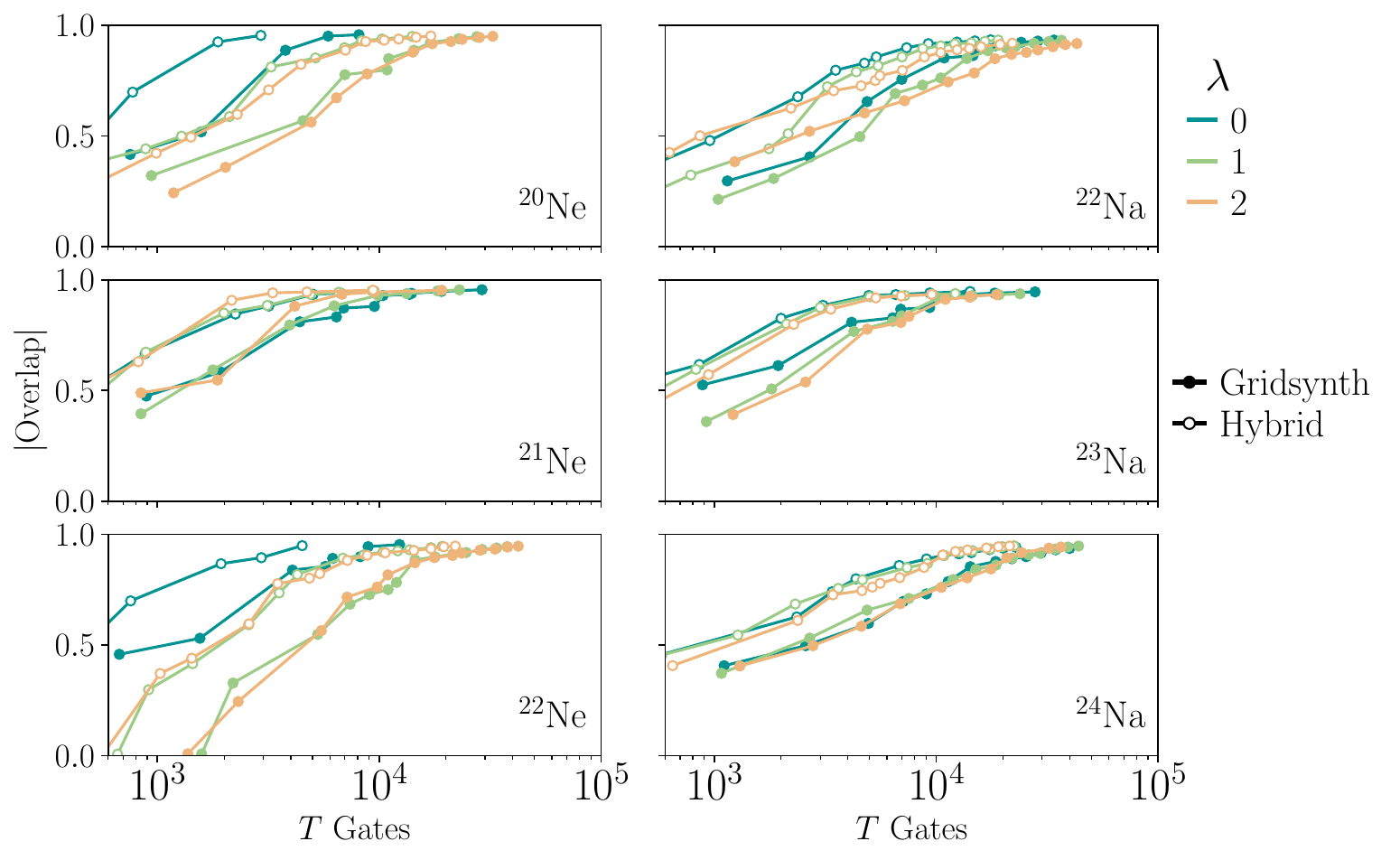}
    \caption{\textbf{Unitary Synthesis.} The optimized circuits described in Sec.~\ref{sec:circuit_opt} are decomposed to the Clifford$+T$ gateset, with the $T$-count of the circuits plotted against the overlap magnitude to the target eigenstates, over the first three eigenstates (numerated by $\lambda \in \{0,1,2\}$) of the nuclear isotopes $^{20-22}$Ne and $^{22-24}$Na. The decomposition using purely $R_z$ rotation synthesis via Gridsynth is shown by the coloured markers; the hybrid decomposition using the direct single-qubit unitary synthesis via Trasyn is shown by the white markers.}
    \label{fig:usdb_approx_TGates}
\end{figure*}


\subsection{Gate Decompositions}\label{sec:decompositions}

Our final goal is generating circuits with a low $T$-count to approximately prepare a target eigenstate. 
We proceed by first expressing the circuit in the Clifford$+R_z$ gateset, and then the $R_z$ gates are further decomposed into the Clifford$+T$ gateset. We aim to reduce the total number of $R_z$ gates, to proportionally reduce the number of $T$ gates to represent these to the target error.
Fortunately, the optimized circuits composed of layers of $SU(4)$ gates permit a simple decomposition to the Clifford$+R_z$ gateset, as each $SU(4)$ unitary can be realized by a circuit of Clifford gates and at most 15 $R_z$ gates~\cite{zhang2003geometric}. This can be parameterized with first having arbitrary single rotations (requiring 3 $R_z$ gates plus Cliffords) on each qubit. This is followed by an $R_{xx}$, $R_{yy}$ and a $R_{zz}$ rotation gate between both qubits, each requiring one $R_z$ gate (plus Cliffords) when decomposed to the Clifford$+R_z$ gateset. Finally another round of arbitrary single qubit rotations on each qubit takes the total number of $R_z$ gates up to 15. See Fig.~\ref{fig:Overview}d) for a visualization.

We note however, that if each $SU(4)$ unitary is expressed in this form, there is an obvious redundancy in the number of $R_z$ gates. 
When consecutive $SU(4)$ unitaries in the circuit are decomposed to the Clifford$+R_z$ gateset, this results in 6 consecutive single qubit rotations on shared qubits.
This is redundant, as any single qubit rotation can be decomposed into at most 3 single qubit rotations. 
This gauge freedom in single-qubit gates allows the extra 3 $R_z$ gates to be compressed away. As a result, other than the final layer of gates, all $SU(4)$ unitaries can be expressed with 9 $R_z$ gates rather than the maximum 15, resulting in a reduction in  $R_z$ gates by $40\%$ for all gate layers other than the last. See Fig.~\ref{fig:redundancy} for a visualisation of this gate redundancy removal.
\medskip

With our circuits now expressed in the Clifford$+R_z$ gateset, we now finally perform the decomposition to the Clifford$+T$ gateset. We first perform the decomposition of the $R_z$ gates with Gridsynth~\cite{ross2014optimal}, an algorithm considered state-of-the-art for this task. For our previously optimized circuits, it is not obvious whether a shallow circuit with reasonable fidelity, or a deeper circuit with high fidelity, will be most favourable for a following Clifford$+T$ decomposition, so this is performed for all circuits. For each Clifford+$R_z$ circuit, we decompose the $R_z$ gates to increasing accuracy, as controlled by the $\epsilon$ input parameter to Gridsynth. For our study, we use a global $\epsilon$, where every $R_z$ gate in the circuit is decomposed to the same precision.
For a given circuit, and global precision value $\epsilon$, we test the infidelity of eigenstate prepared by the approximate circuits. Smaller values of $\epsilon$ results in the more accurate approximation of each $R_z$ gate, and therefore introduces smaller errors in the preparation of the eigenstate at the expense of a higher $T$-count. For each circuit, we perform this decomposition across the range of $\epsilon \in \{10^{-1.0}, 10^{-1.5},10^{-2.0},10^{-2.5}\}$, producing a dataset of circuits with a total number of $T$ gates, and corresponding infidelity with the target eigenstate. 

The structure of our circuits can be further exploited for gains in the decomposition to the Clifford$+T$ gateset. Gridsynth is highly optimized for decompositions of $R_z$ gates by exploiting simplifications due to the diagonal gate structure. However for many instances in our circuits, on a particular qubit there are 3 $R_z$ gates interspersed with single qubit Cliffords. Therefore, methods directly decomposing single qubit unitaries may provide a significant advantage, compared to individually decomposing each of the 3 $R_z$ gates separately with Gridsynth.

For this purpose we use Trasyn ~\cite{hao2025reducing}, a recent algorithm proposed for unitary synthesis of $U3$ gates. Here the unitary synthesis is treated as a search problem, where we seek a sequence of Clifford gates and few $T$ gates applied in sequence, that maximizes a unitary fidelity with the target unitary. Exhaustively searching this search space is exponentially costly in the number of $T$ gates in the circuit, Trasyn overcomes this by expressing the search space by an MPS chain and using tensor network sampling methods to search for Clifford$+T$ gate sequences best approximating the target unitary. They found up to a $3\times$ reduction in $T$ gates required to approximate single qubit unitaries, compared to decomposing the 3 $R_z$ gates (plus Clifford gates) individually with Gridsynth. Given that in our circuit construction, the 2-qubit gates deriving from the original $SU(4)$ unitaries, these can be expressed by 9 $R_z$ gates (apart from the last layer which is 15), and 6 of these appear in sequences of 3 $R_z$ gates (plus Clifford), there is the potential to reduce to the effective cost of 5 $R_z$ gates, or a further reduction by 44.4\%. See Fig.~\ref{fig:synthesis_methods} for a visualisation of the two decomposition methods.
As done for the purely $R_z$ synthesis, we repeat the same procedure of iterating over circuit depths and $\epsilon$ values to perform the unitary synthesis to, however now any time there are 3 consecutive $R_z$ gates (plus single qubit Clifford gates) we instead decompose these with Trasyn.

In Fig.~\ref{fig:usdb_approx_TGates}, we show the final results of our circuit compilation to a fault-tolerant gateset. We compare the two methods of gate decomposition, with only $R_z$ decompositions using Gridsynth, and the mixed use of $R_z$ and $U3$ decompositions with Gridsynth and Trasyn, denoted by the filled and unfilled markers respectively. Each point on the plot is a circuit with a total number of $T$ gates, given by its position along the $x$-axis, and the infidelity it prepares the target state to, given by its position along the $y$-axis. For a particular decomposition method and nuclear eigenstate, we iterate over a range of circuit depths and $\epsilon$ value before applying the unitary synthesis, resulting in a dataset; we filter these to show the Pareto front of best performing circuits.
Across all nuclei tested, we find the eigenstates can be well approximated with low $T$-counts, with $\sim10^4$ $T$ gates sufficient to prepare the eigenstates to an overlap magnitude of $\sim0.8$.

\begin{figure*}[t]
    \centering
    \includegraphics[width=0.9\linewidth]{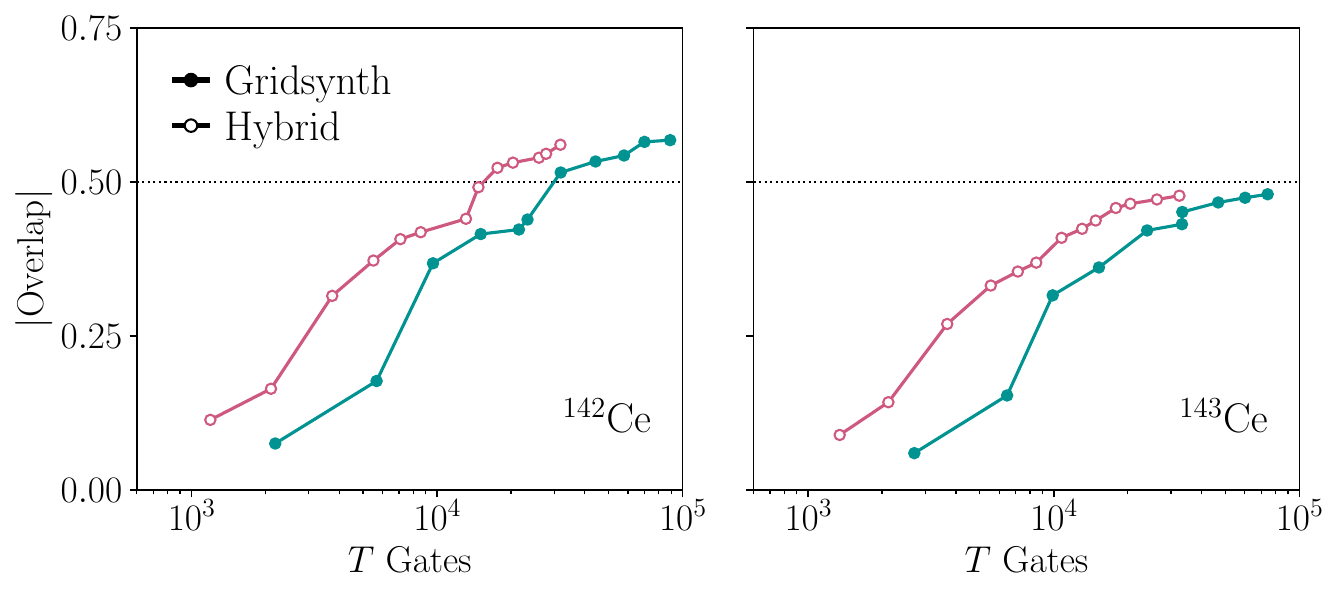}
    \caption{\textbf{Large Circuit Synthesis.} Using the circuit optimization and synthesis methods developed, we give the final overlaps of circuits preparing the ground states of the $^{142}$Ce and $^{143}$Ce isotopes, where the Hamiltonian is represented by 76 qubits. 
    The total number of $T$ gates in the circuit ($U$) versus the overlap magnitude with the target ground state is plotted. We stress this is the estimated  lower bound of the fidelity with the true ground state using the inequality Eq.~\eqref{eq:lower_bound}, using the exactly computed overlap $\big|\langle\Phi(256)|U|\mathbf{0}\rangle\big|$ and the estimated overlap $\big|\langle\Phi(256)|\Phi(\infty)\rangle\big|$.
    To capture high overlap magnitudes of 0.5 (highlighted by the horizontal dotted line), we find only $\sim2\times 10^4$ $T$ gates are required. }
    \label{fig:Ce_synthesis}
\end{figure*}

\medskip

\newpage

\section{Large Scale Implementation} \label{sec:final_test}

In the previous sections we described the methods developed for compiling circuits approximating eigenstates, and synthesized to the Clifford$+T$ gateset.
We now apply these methods to larger systems beyond the limits of precise study with DMRG.

\medskip

For these larger scale demonstrations, we target the nuclear ground states of the cerium isotopes, $^{142}$Ce and $^{143}$Ce.
The valence space used has 76 single particle orbitals, 32 proton and 44 neutron. These ground states exist in the symmetry sector with 8 protons and 2 (3) neutrons for the  $^{142}$Ce ($^{143}$Ce) isotope. These correspond to subspace dimensions of $\binom{32}{8}\times\binom{44}{2}=9.95\times 10^9$ and $\binom{32}{8}\times\binom{44}{3}=1.39\times 10^{11}$ respectively.
See Section~\ref{sec:hamiltonian} for the details on form of this Hamiltonian and valence space.

\medskip

The Hamiltonian required a large bond dimension to express as an MPO, at a maximum of 1191. The construction of this MPO was greatly accelerated using the \texttt{ITensorMPOConstruction} package \cite{corbett2025scaling}.
We apply DMRG to this Hamiltonian, using a variational MPS with the largest accessible bond dimension that could fit the calculation within the 80Gb memory available on an Nvidia H100 GPU card, which was $\chi=1000$. 
Ultimately we care about the overlap of this MPS approximation with the exact ground state. While this is not exactly computable, there are a range of methods that could be used to estimate this. We follow the extrapolation procedure introduced in~\cite{berry2025rapid}, with the technical details given in our Appendix.~\ref{sec:extrapolation}. A reverse-sweep DMRG procedure is used to create the dataset of MPS for this extrapolation; here the DMRG optimization is run at the largest accessible maximum bond dimension (with a random initalization) until convergence, then we iteratively decrease the maximum bond dimension with the previously outputted MPS as an initialization.
The extrapolation estimates the largest bond dimension MPS to have an overlap magnitude to the true ground state of $\big|\langle\Phi(\chi=1000)|\Phi(\chi=\infty) \rangle|$ equal to 0.985 (0.979) for $^{142}$Ce ($^{143}$Ce). This highlights that while MPS are effective at approximating these nuclear systems, the ground states are beyond exact study with large-scale DMRG.

\medskip

We then proceed to the variational compilation of a circuit to approximate the ground state. 
Due to the circuit compilation code not exploiting GPU acceleration, to aid the algorithm tractability we use a lower bond dimension state with $\chi=256$ again produced by a reverse-sweep DMRG. By the extrapolation described in the Appendix~\ref{sec:extrapolation}
and shown in Fig.~\ref{fig:fidelity_extrapolate}, this lower bond dimension state is still estimated to have a large overlap of $|\langle\Phi(256)|\Phi(\infty)\rangle| = $ 0.960 (0.943) for $^{142}$Ce  ($^{143}$Ce).
For increasing depth circuits, the iterative optimization produces circuits with 1 to 6 staircase layers (with 75 SU(4) unitaries in each layer),  that when applied to the $|\mathbf{0}\rangle$ state have overlap magnitudes ($|\langle\Psi(256)|U|\mathbf{0}\rangle|$) with the target MPS of 
$[0.281, 0.622, 0.720, 0.786, 0.801, 0.837]$ for the $^{142}$Ce ground state, and $[0.272, 0.601, 0.707, 0.771, 0.786, 0.792]$ for $^{143}$Ce.


In this compilation, there are three states to consider:
\begin{itemize}
    \item $|\psi(\boldsymbol{\theta})\rangle = U(\boldsymbol{\theta})|\mathbf{0}\rangle$, the state generated by the compiled circuit
    \item $|\Phi(\chi=256)\rangle$, the target MPS produced by the DMRG calculation
    \item $|\Phi(\chi=\infty)\rangle$, the exact ground state in the limit of infinite bond dimension.
\end{itemize}

We can directly compute $|\langle\Psi(256)|\psi(\boldsymbol{\theta})\rangle|$, and we estimate $|\langle\Psi(256)|\Psi(\infty)\rangle|$ using the extrapolation procedure detailed in Appendix ~\ref{sec:extrapolation}. Using these two values, we ultimately we want to estimate the quantity $|\langle\psi(\boldsymbol{\theta})|\Phi(\infty)\rangle|$, and in the Appendix~\ref{sec:triangle_ineq}, we derive the following lower-bound using a triangle inequality on the Wootters' distance between states

\begin{equation}\label{eq:lower_bound}
\begin{split}
    |\langle\psi(\boldsymbol{\theta})|\Phi(\infty)\rangle| \geq &|\langle\psi(\boldsymbol{\theta})|\Phi(256)\rangle|
    \times |\langle\Phi(256)|\Phi(\infty)\rangle| \\
    &-\sqrt{1-|\langle\psi(\boldsymbol{\theta})|\Phi(256)\rangle|^2} \\
    &\times \sqrt{1-|\langle\Phi(256)|\Phi(\infty)\rangle|^2}. \\
\end{split}
\end{equation}

\medskip

Finally, we take the low-depth circuits composed of layers of $SU(4)$ unitaries, and synthesize the rotation gates to the Clifford + $T$ gateset. After each 2-qubit unitary is decomposed by the KAK decomposition, we remove the redundant single-qubit rotations between the 2-qubit rotations, as described in Sec~\ref{sec:decompositions}, and visually shown in Fig.~\ref{fig:redundancy}. Then we use the same two synthesis procedures: the first using only Gridsynth to decompose every $R_z$ gate to the Clifford$+T$ gateset; the second uses the hybrid strategy which identifies all instances where there are more general single qubit rotations requiring 3 $R_z$ gates, which are compiled directly with Trasyn, and the isolated $R_z$ gates are compiled with Gridsynth. Again, we iterate over each depth circuit and decompose every gate to the same precision, $\epsilon \in [10^{-1}, 10^{-1.5}, 10^{-2}, 10^{-2.5}]$. We show this data in Fig.~\ref{fig:Ce_synthesis}, where over the range of circuit depths and synthesis precisions, we plot the Pareto front of highest fidelity circuits for a given number of total $T$ gates. We stress that the plotted fidelity is the estimated lower-bound on the overlap magnitude to the exact ground state, given by the inequality in Eq.~\eqref{eq:lower_bound}. 


The hybrid strategy again outperforms the pure Gridsynth decomposition strategy, and for these ground states expressed on 76 qubits, we estimate the exact ground state can be prepared to a high overlap magnitude of 0.5 with circuits of ~$\sim 2\times 10^4$ $T$ gates.

\section{Discussion}



We have presented an efficient protocol for preparing nuclear shell model eigenstates on fault-tolerant quantum computers by combining tensor network methods, variational quantum circuit compilation, and unitary synthesis. Across all systems tested --- from 24-qubit $sd$-shell nuclei verifiable by exact diagonalization, to 76-qubit cerium isotopes with Hilbert space dimensions exceeding $10^{10}$ --- we consistently find circuits requiring only up to $\sim 2 \times 10^4$ total $T$ gates to prepare the target eigenstates to high fidelity. 
The combination of these new compilation techniques, with a focused study applied to Nuclear Structure, found low resource requirements to prepare approximations of these ground states, and provides a new optimistic baseline for Initial State Preparation on early fault-tolerant quantum computers.
Ultimately, these results should be combined with resource estimates for the energy estimation routines to understand the end-to-end overheads for the full QPE algorithm.
While fault-tolerant resource estimates of QPE for nuclear structure have received little study compared to quantum chemistry, a recent study targetting pionless EFT Hamiltonians indicates interesting nuclear dynamics could be simulated with upwards of tens of millions of $T$ gates~\cite{spagnoli2025quantum}.

\subsection*{The classical-quantum crossover}

A fundamental question arising from our reliance on DMRG is whether a quantum computer remains necessary if the state can be approximated classically. The answer lies in distinguishing \emph{sufficient fidelity} from \emph{numerical precision}. QPE does not require an exact input state; it requires only a guide state with high overlap, as the sampling overheads scaling inversely with the overlap magnitude under amplitude amplification~\cite{ge2019faster}. Our results demonstrate that DMRG can efficiently provide this sufficient overlap even when exact classical representation is intractable.

The cerium isotope calculations illustrate this concretely. For $^{142}$Ce and $^{143}$Ce, with Hilbert space dimensions of approximately $10^{10}$ and $10^{11}$, an MPS ($\chi = 256$) achieved estimated squared overlaps of 0.960 and 0.943 with the true ground states --- high fidelity, but not exact representation. Added approximations due to the subsequent circuit compilation and Clifford+$T$ synthesis reduced this further, yet the estimated lower bound on the final overlap magnitude with the true ground state is still high, resulting in a minimal sampling overhead in the target QPE simulation. This establishes a symbiotic \emph{crossover regime}: classical tensor network computation provides a coarse approximation, and the quantum computer performs precision spectroscopy via phase estimation. 

\subsection*{Sources of resource efficiency}

The low $T$-counts arise from three synergistic elements. First, the physics-aware orbital mapping --- segregating proton and neutron sectors and ordering orbitals to place strongly paired states on neighboring MPS sites --- capitalizes on the known entanglement structure of nuclear ground states~\cite{perez2023quantum, brokemeier2025quantum, johnson2023proton}, suppressing the bond dimensions required by DMRG and the circuit depth of the subsequent compilation.

Second, the variational circuit compilation trades exactness for the flexibility of a shallow circuit ansatz of general $\mathrm{SU}(4)$ unitaries.
This allows the direct targetting of circuit structures requiring few $R_z$ gates, reducing the number of rotation gates that require synthesizing to the Clifford$+T$ gateset.

Third, the gate decomposition strategy exploits the specific structure of our circuits. Removing redundant single-qubit rotations between consecutive $\mathrm{SU}(4)$ blocks reduces the $R_z$ count by 40\% (Fig.~\ref{fig:redundancy}). The hybrid use of Trasyn~\cite{hao2025reducing} for direct $U_3$ synthesis, compiling three consecutive $R_z$ gates as a single unitary rather than individually with Gridsynth~\cite{ross2014optimal}, provides a further reduction of up to 44\%. This hybrid strategy applies to any circuit compiled from layers of $\mathrm{SU}(4)$ unitaries, independent of the physical application.

\subsection*{Scalability, limitations, and outlook}

Our $sd$-shell benchmarks (24 qubits) allow exact verification of all fidelities and energies. The larger calculations (76 qubits) in Section ~\ref{sec:final_test} extend the method to the frontier of classical simulation, where extrapolation procedures become necessary. That accurate results are achievable in this intermediate regime, practically relevant yet amenable to classical cross-checks, is a strength of the current study.

Extensions to heavier nuclei with stronger proton-neutron correlations or deformed structure may weaken the favourable entanglement properties exploited here, requiring larger MPS bond dimensions. The linear MPS topology may also limit the capture of multi-scale correlations. Promising directions include alternative tensor network ans\"{a}tze~\cite{orus2019tensor} such as tree tensor networks or MERA, which can more naturally represent higher-dimensional correlation structures. On the circuit side, problem-inspired ans\"{a}tze tailored to nuclear symmetries~\cite{sarma_low-circuit-depth_2026, bhoy2024shell} or adaptive circuit constructions~\cite{grimsley2019adaptive, perez2023nuclear, jaderberg2025variational} could further reduce $R_z$ counts and hence final $T$-counts. 

More broadly, the dual use of tensor networks, for classical eigenstate approximation via DMRG and for unitary synthesis via Trasyn, provides a modular framework where each component can be independently improved through advances in GPU-accelerated DMRG~\cite{hyatt2019dmrg, ganahl2023density, menczer2024two, brower2025mixed} or unitary synthesis algorithms. The framework applies to any fermionic system where tensor network methods capture sufficient fidelity; quantum chemistry~\cite{chan2008introduction} is a natural next target. For systems beyond the reach of DMRG, complementary methods such as sparse Pauli propagation~\cite{rudolph2025pauli, miller2025simulation, lin2025utility} could provide alternative initial states within the same compilation framework. 

Looking forward, the most pressing need is end-to-end resource estimation combining the Initial State Preparation costs quantified here with concrete estimates for the time-evolution operator and phase estimation arithmetic, providing a roadmap toward quantum advantage in nuclear structure physics.


\begin{acknowledgments}
CS acknowledges support from the UK STFC under grant ST/Y000358/1. LC acknowledges support by the Laboratory Directed Research and Development (LDRD) program of Los Alamos National Laboratory (LANL) under project number 20260043DR and by the U.S. Department of Energy, Office of Science, Office of Advanced Scientific Computing Research under Contract No. DE-AC05-00OR22725 through the Accelerated Research in Quantum Computing Program MACH-Q project. UK Ministry of Defence $\copyright$Crown owned copyright 2026/AWE.
\end{acknowledgments}


\newpage

\bibliography{quantum.bib,extra}

\clearpage 
\input{Supplementary}


\end{document}

%% file: supplementary.tex


\appendix

\setcounter{page}{1}
\def\theHpage{A\arabic{page}}
\def\theHequation{A\arabic{equation}}

\setcounter{figure}{0} 
\renewcommand{\thefigure}{A\arabic{figure}}
\renewcommand{\theHfigure}{A\arabic{figure}}

\setcounter{equation}{0}
\renewcommand{\theequation}{A\arabic{equation}}
\renewcommand{\theHequation}{A\arabic{equation}}

\onecolumngrid

\begin{center}
\large{ Supplementary Material for \\ `` Low $T$-count preparation of nuclear eigenstates with tensor networks ''
}
\end{center}


\section{Overlap Extrapolation to Exact Ground State}\label{sec:extrapolation}

In Sec.~\ref{sec:final_test}, we work with nuclear Hamiltonians with a ground state beyond exact representation by MPS, and therefore we do not have access to the true ground state wavefunction. To estimate the overlap of MPS representations to the ground state, we use the extrapolation procedure introduced in ~\cite{berry2025rapid}.
Given a series of MPS produced by DMRG over a range of maximum bond dimensions, $\chi$, the following linear relations were observed

\begin{equation}\label{eq:linear1}
    \text{log}(1 - \big|\langle \Phi(\chi')|\Phi(\infty) \rangle\big|^2) \text{ vs. } (\text{log}(\chi'))^2 
\end{equation}
\begin{equation}\label{eq:linear2}
\begin{split}
    & \log(\big|\langle \Phi(\chi')|\Phi(\chi'') \rangle\big|^2 - \big|\langle \Phi(\chi')|\Phi(\infty) \rangle\big|^2 ) \text{ vs. } (\log(\chi''))^2 \\
    & \qquad \text{where } \chi' \ll \chi''
\end{split}
\end{equation}
We use these relations to enable an extrapolation to an estimate for the quantity $\big|\langle\Phi(\chi')\Phi(\infty)|^2$.
We first apply a reverse-sweep DMRG procedure to produce a series of MPS approximations. 
Upon convergence of this DMRG optimization starting at a bond dimension of $\chi=1000$, we iteratively decreased the maximum bond dimension, using the previously converged MPS as an initialisation. This reverse-sweep procedure created a dataset  $\{|\Phi(\chi_j)\}_j$  of MPS approximations of the ground state with $\chi'' \in \{600,700,800,900,1000\}$ and $\chi' \in \{20,40,60,80,100\}$.

\begin{figure*}[b]
    \centering
    \includegraphics[width=0.8\linewidth]{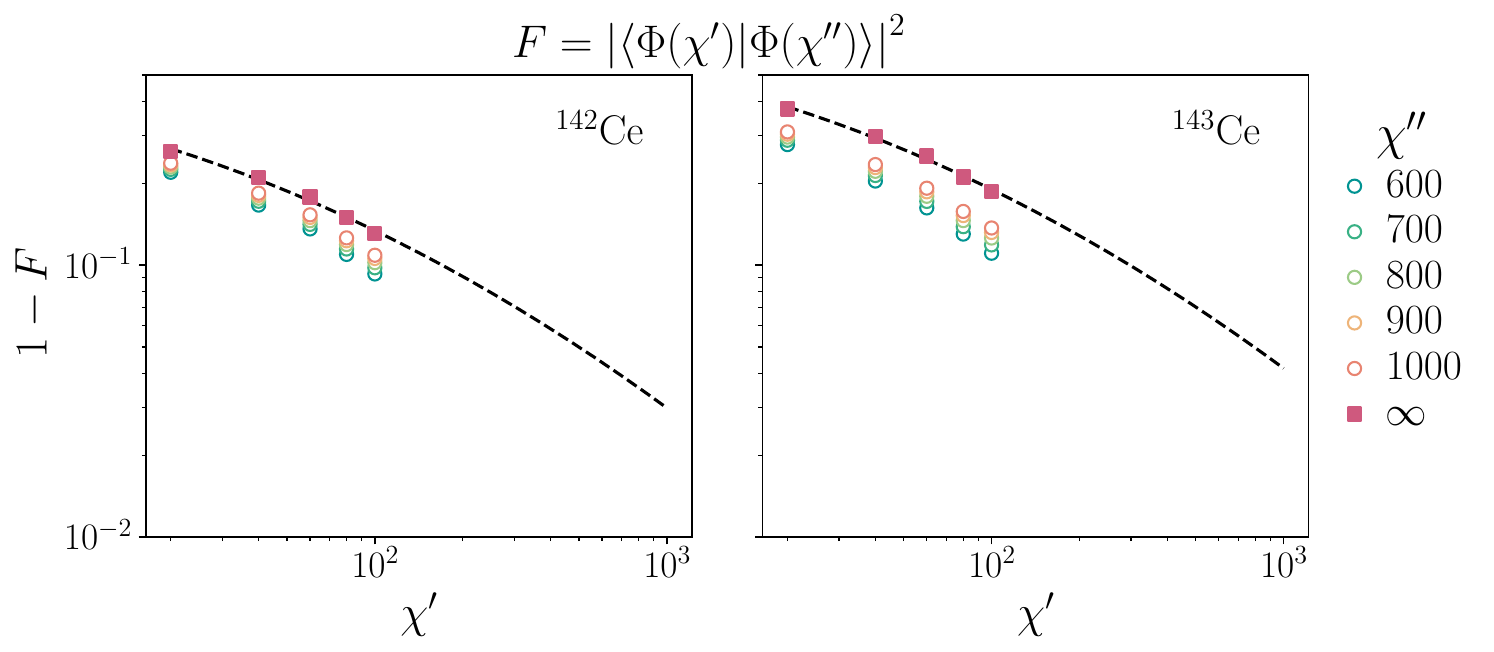}
    \caption{\textbf{Fidelity Extrapolation.} Data showing the extrapolation to estimate the overlap of finite bond dimension MPS with the exact ground state. First, we compute $1-F = 1-\big|\langle\Phi(\chi')|\Phi(\chi'')\rangle\big|^2$ over the range of $\chi'$ and $\chi''$ values, shown by the coloured circles. Then we use the relation in Eq.~\eqref{eq:linear2} to extrapolate and estimate the quantity $1-\big|\langle\Phi(\chi')|\Phi(\infty)\rangle\big|^2$ for the small $\chi'$ values, shown by the red squares. Then, this data is used to perform another fit, based on Eq.~\eqref{eq:linear1}, where the dotted line shows the quadratic fit between $\text{log}(1-F)$ and $\text{log}(\chi')$. This allows an estimate for $1-\big|\langle\Phi(\chi')|\Phi(\infty)\rangle\big|^2$ for all $\chi'$ values.}
    \label{fig:fidelity_extrapolate}
\end{figure*}

Using this dataset with $\chi' \ll \chi''$, we first perform a fit using the relation Eq.~\eqref{eq:linear2} to estimate the quantity $\big|\langle \Phi(\chi')|\Phi(\infty) \rangle\big|^2$. The computed datapoints are shown by the coloured circles with white faces in Fig.~\ref{fig:fidelity_extrapolate}, and the extrapolated datapoints are shown by the red squares.

Then, these values are used in Eq.~\eqref{eq:linear1} to extrapolate the squared overlap $\big|\langle \Phi(\chi')|\Phi(\infty) \rangle\big|^2$ for  larger $\chi'$ values, with the extrapolation shown by the black dashed line in Fig.~\ref{fig:fidelity_extrapolate}. This estimates the maximum bond dimension MPS computed $|\Phi(1000)\rangle$ having a squared  overlap to the exact ground state of $\big|\langle \Phi(1000)|\Phi(\infty) \rangle\big|^2$ = 0.970 (0.958) for the $^{142}$Ce ($^{143}$Ce) ground state.

\clearpage


\section{Triangle Inequality}\label{sec:triangle_ineq}

Given three normalized states $|A\rangle$, $|B\rangle$ and $|C\rangle$, where we can compute $\left|\langle A| B\rangle\right|$ and $\left|\langle B| C\rangle\right|$, we want to use these values to bound $\left|\langle A| C\rangle\right|$.

\medskip

The appropriate quantity used to bound the distance between these states is the Wootters' distance~\cite{wootters1981statistical}, $$d_\text{W}(|A\rangle,|B\rangle) = \text{arcos}\left(| \langle A|B \rangle|\right).$$ 

The metric character of the Wootters' distance gives the following triangle inequality ~\cite{bosyk2014geometric}

\begin{equation}
d_\text{W}(|A\rangle,|C\rangle) \leq d_\text{W}(|A\rangle,|B\rangle) + d_\text{W}(|B\rangle,|C\rangle).
\end{equation}

Let $\theta_1 = d_\text{W}(|A\rangle,|B\rangle)$, $\theta_2 = d_\text{W}(|B\rangle,|C\rangle)$. Then

\begin{equation}
    \begin{split}
        \text{cos}(d_\text{W}(|A\rangle,|C\rangle)) \geq & \ \text{cos}(\theta_1 + \theta_2) \\
        &= \text{cos}(\theta_1)\text{cos}(\theta_2) - \text{sin}(\theta_1)\text{sin}(\theta_2).
    \end{split}
\end{equation}

Using $\text{cos}(d_\text{W}(|A\rangle,|B\rangle))=\left| \langle A|B \rangle \right|$ and $\text{sin}(d_\text{W}(|A\rangle,|B\rangle))=\text{sin}(\text{arcos}(\left| \langle A|B \rangle \right|))= \sqrt{1-\left| \langle A|B \rangle \right|^2}$

\medskip

This results in the lower bound relating the three overlap magnitudes

\begin{equation}
    \begin{split}
        \left| \langle A|C \rangle \right| \geq  \left| \langle A|B \rangle \right| \left| \langle B|C \rangle \right| - \sqrt{1-| \langle A|B \rangle |^2}\sqrt{1-| \langle B|C \rangle |^2}.
    \end{split}
\end{equation}

%% file: extra.bib
@book{deshalittalmi,
    title={Nuclear {S}hell {T}heory},
    author={de-{S}halit, Amos and Talmi, Igal},
    year = {1963},
    publisher = {Academic Press, New York and London}}

@article{RevModPhys.75.607,
  title = {Pairing in nuclear systems: from neutron stars to finite nuclei},
  author = {Dean, D. J. and Hjorth-Jensen, M.},
  journal = {Rev. Mod. Phys.},
  volume = {75},
  issue = {2},
  pages = {607--656},
  numpages = {0},
  year = {2003},
  month = {Apr},
  publisher = {American Physical Society},
  doi = {10.1103/RevModPhys.75.607},
  url = {https://link.aps.org/doi/10.1103/RevModPhys.75.607}
}

@article{PhysRevC.74.034315,
  title = {New ``USD'' Hamiltonians for the $\mathit{sd}$ shell},
  author = {Brown, B. Alex and Richter, W. A.},
  journal = {Phys. Rev. C},
  volume = {74},
  issue = {3},
  pages = {034315},
  numpages = {11},
  year = {2006},
  month = {Sep},
  publisher = {American Physical Society},
  doi = {10.1103/PhysRevC.74.034315},
  url = {https://link.aps.org/doi/10.1103/PhysRevC.74.034315}
}

@article{PhysRevC.44.233,
  title = {First-forbidden \ensuremath{\beta} decay in the lead region and mesonic enhancement of the weak axial current},
  author = {Warburton, E. K.},
  journal = {Phys. Rev. C},
  volume = {44},
  issue = {1},
  pages = {233--260},
  numpages = {0},
  year = {1991},
  month = {Jul},
  publisher = {American Physical Society},
  doi = {10.1103/PhysRevC.44.233},
  url = {https://link.aps.org/doi/10.1103/PhysRevC.44.233}
}

@article{sarma_low-circuit-depth_2026,
	title = {A low-circuit-depth quantum computing approach to the nuclear shell model},
	volume = {2},
	issn = {3059-4529},
	url = {https://doi.org/10.1007/s44464-026-00009-9},
	doi = {10.1007/s44464-026-00009-9},
	number = {1},
	urldate = {2026-02-17},
	journal = {Discover Quantum Science},
	author = {Sarma, Chandan and Stevenson, P. D.},
	month = feb,
	year = {2026},
	pages = {6},
}


%% file: quantum.bib
@article{schollwock2011density,
  title={The density-matrix renormalization group in the age of matrix product states},
  author={Schollw{\"o}ck, Ulrich},
  journal={Annals of physics},
  volume={326},
  number={1},
  pages={96--192},
  year={2011},
  publisher={Elsevier},
  url={https://www.sciencedirect.com/science/article/pii/S0003491610001752?via%3Dihub},
  url={10.1016/j.aop.2010.09.012}
}

@article{corbett2025scaling,
  title={Scaling up the transcorrelated density matrix renormalization group},
  author={Corbett, Benjamin and Miyake, Akimasa},
  journal={Physical Review B},
  volume={112},
  number={16},
  pages={165120},
  year={2025},
  publisher={APS},
  url={https://journals.aps.org/prb/abstract/10.1103/nzrt-l2j1},
  doi={https://doi.org/10.1103/nzrt-l2j1}
}

@article{ballarin2025efficient,
  title={Efficient quantum state preparation of multivariate functions using tensor networks},
  author={Ballarin, Marco and Garc{\'\i}a-Ripoll, Juan Jos{\'e} and Hayes, David and Lubasch, Michael},
  journal={arXiv preprint arXiv:2511.15674},
  year={2025},
  url={https://arxiv.org/abs/2511.15674},
  doi={10.48550/arXiv.2511.15674}
}

@article{rudolph2022decomposition,
  title={Decomposition of matrix product states into shallow quantum circuits},
  author={Rudolph, Manuel S and Chen, Jing and Miller, Jacob and Acharya, Atithi and Perdomo-Ortiz, Alejandro},
  journal={Quantum Science and Technology},
  volume={9},
  number={1},
  pages={015012},
  year={2023},
  publisher={IOP Publishing},
  url={https://iopscience.iop.org/article/10.1088/2058-9565/ad04e6},
}

@article{white1992density,
  title={Density matrix formulation for quantum renormalization groups},
  author={White, Steven R},
  journal={Physical Review Letters},
  volume={69},
  number={19},
  pages={2863},
  year={1992},
  publisher={APS},
  doi={10.1103/PhysRevLett.69.2863}
}

@article{grimsley2019adaptive,
  title={An adaptive variational algorithm for exact molecular simulations on a quantum computer},
  author={Grimsley, Harper R and Economou, Sophia E and Barnes, Edwin and Mayhall, Nicholas J},
  journal={Nature {C}ommunications},
  volume={10},
  number={1},
  pages={1--9},
  year={2019},
  publisher={Nature Publishing Group},
  url={https://www.nature.com/articles/s41467-019-10988-2},
  doi={10.1038/s41467-019-10988-2}
}

@article{berry2025rapid,
  title={Rapid Initial-State Preparation for the Quantum Simulation of Strongly Correlated Molecules},
  author={Berry, Dominic W and Tong, Yu and Khattar, Tanuj and White, Alec and Kim, Tae In and Low, Guang Hao and Boixo, Sergio and Ding, Zhiyan and Lin, Lin and Lee, Seunghoon and others},
  journal={PRX Quantum},
  volume={6},
  number={2},
  pages={020327},
  year={2025},
  publisher={APS},
  url={https://journals.aps.org/prxquantum/abstract/10.1103/PRXQuantum.6.020327},
  doi={10.1103/PRXQuantum.6.020327}
}

@article{weinberg2017quspin,
  title={QuSpin: a Python package for dynamics and exact diagonalisation of quantum many body systems part I: spin chains},
  author={Weinberg, Phillip and Bukov, Marin},
  journal={SciPost Physics},
  volume={2},
  number={1},
  pages={003},
  year={2017},
  url={https://www.scipost.org/SciPostPhys.2.1.003},
  doi={10.21468/SciPostPhys.2.1.003}
}

@article{carrasco2025comparison,
  title={Comparison of variational quantum eigensolvers in light nuclei},
  author={Carrasco-Codina, Miquel and Costa, Emanuele and Romero, Antonio M{\'a}rquez and Men{\'e}ndez, Javier and Rios, Arnau},
  journal={arXiv preprint arXiv:2507.13819},
  year={2025},
  url={https://arxiv.org/abs/2507.13819},
  doi={10.48550/arXiv.2507.13819}
}

@article{yoshida2025bridging,
  title={Bridging quantum computing and nuclear structure: Atomic nuclei on a trapped-ion quantum computer},
  author={Yoshida, Sota and Sato, Takeshi and Ogata, Takumi and Kimura, Masaaki},
  journal={Physical Review Research},
  volume={8},
  number={1},
  pages={013134},
  year={2026},
  publisher={APS},
  url={https://journals.aps.org/prresearch/abstract/10.1103/td9s-z7my},
  doi={10.1103/td9s-z7my}
}

@article{costa2025quantum,
  title={A quantum annealing protocol to solve the nuclear shell model},
  author={Costa, Emanuele and Perez-Obiol, Axel and Menendez, Javier and Rios, Arnau and Garcia-Saez, Artur and Juli{\'a}-D{\'\i}az, Bruno},
  journal={SciPost Physics},
  volume={19},
  number={2},
  pages={062},
  year={2025},
  url={https://www.scipost.org/10.21468/SciPostPhys.19.2.062},
  doi={10.21468/SciPostPhys.19.2.062}
}

@article{costa2025quasiparticle,
  title={Quasiparticle pairing encoding of atomic nuclei for quantum annealing},
  author={Costa, Emanuele and P{\'e}rez-Obiol, Axel and Men{\'e}ndez, Javier and Rios, Arnau and Garc{\'\i}a-S{\'a}ez, Artur and Juli{\'a}-D{\'\i}az, Bruno},
  journal={arXiv preprint arXiv:2510.10118},
  year={2025},
  url={https://arxiv.org/abs/2510.10118},
  doi={10.48550/arXiv.2510.10118}
}

@article{spagnoli2025quantum,
  title={Quantum Simulation of Nuclear Dynamics in First Quantization},
  author={Spagnoli, Luca and Lissoni, Chiara and Roggero, Alessandro},
  journal={arXiv preprint arXiv:2507.22814},
  year={2025},
  url={https://arxiv.org/abs/2507.22814},
  doi={10.48550/arXiv.2507.22814}
}

@article{dukelsky2001new,
  title={New approach to large-scale nuclear structure calculations},
  author={Dukelsky, Jorge and Pittel, S},
  journal={Physical Review C},
  volume={63},
  number={6},
  pages={061303},
  year={2001},
  publisher={APS},
  url={https://journals.aps.org/prc/abstract/10.1103/PhysRevC.63.061303},
  doi={10.1103/PhysRevC.63.061303}
}

@article{pittel2006density,
  title={Density matrix renormalization group and the nuclear shell model},
  author={Pittel, S and Sandulescu, N},
  journal={Physical Review C—Nuclear Physics},
  volume={73},
  number={1},
  pages={014301},
  year={2006},
  publisher={APS},
  url={https://journals.aps.org/prc/abstract/10.1103/PhysRevC.73.014301},
  doi={10.1103/PhysRevC.73.014301}
}

@article{poulin2018quantum,
  title={Quantum algorithm for spectral measurement with a lower gate count},
  author={Poulin, David and Kitaev, Alexei and Steiger, Damian S and Hastings, Matthew B and Troyer, Matthias},
  journal={Physical review letters},
  volume={121},
  number={1},
  pages={010501},
  year={2018},
  publisher={APS},
  url={https://journals.aps.org/prl/abstract/10.1103/PhysRevLett.121.010501},
  doi={10.1103/PhysRevLett.121.010501}
}

@article{stoudenmire2012studying,
  title={Studying two-dimensional systems with the density matrix renormalization group},
  author={Stoudenmire, Edwin M and White, Steven R},
  journal={Annu. Rev. Condens. Matter Phys.},
  volume={3},
  number={1},
  pages={111--128},
  year={2012},
  publisher={Annual Reviews},
  url={https://www.annualreviews.org/content/journals/10.1146/annurev-conmatphys-020911-125018},
  doi={10.1146/annurev-conmatphys-020911-125018}
}

@article{orus2014practical,
  title={A practical introduction to tensor networks: Matrix product states and projected entangled pair states},
  author={Or{\'u}s, Rom{\'a}n},
  journal={Annals of Physics},
  volume={349},
  pages={117--158},
  year={2014},
  publisher={Elsevier},
  doi = {10.1016/j.aop.2014.06.013}
}

@article{perez2023nuclear,
  title={Nuclear shell-model simulation in digital quantum computers},
  author={P{\'e}rez-Obiol, Axel and Romero, AM and Men{\'e}ndez, J and Rios, A and Garc{\'\i}a-S{\'a}ez, A and Juli{\'a}-D{\'\i}az, B},
  journal={Scientific Reports},
  volume={13},
  number={1},
  pages={12291},
  year={2023},
  publisher={Nature Publishing Group UK London},
  url={https://www.nature.com/articles/s41598-023-39263-7},
  doi={10.1038/s41598-023-39263-7}
}

@article{hyatt2019dmrg,
  title={$\text{DMRG}$ approach to optimizing two-dimensional tensor networks},
  author={Hyatt, Katharine and Stoudenmire, E Miles},
  journal={arXiv preprint arXiv:1908.08833},
  year={2019},
  url={https://arxiv.org/abs/1908.08833},
  doi={10.48550/arXiv.1908.08833}
}

@article{menczer2024two,
  title={Two-dimensional quantum lattice models via mode optimized hybrid CPU-GPU density matrix renormalization group method},
  author={Menczer, Andor and Kap{\'a}s, Korn{\'e}l and Werner, Mikl{\'o}s Antal and Legeza, {\"O}rs},
  journal={Physical Review B},
  volume={109},
  number={19},
  pages={195148},
  year={2024},
  publisher={APS},
  url={https://journals.aps.org/prb/abstract/10.1103/PhysRevB.109.195148},
  doi={10.1103/PhysRevB.109.195148}
}

@article{brower2025mixed,
  title={Mixed-precision ab initio tensor network state methods adapted for NVIDIA Blackwell technology via emulated FP64 arithmetic},
  author={Brower, Cole and Bernabeu, Samuel Rodriguez and Hammond, Jeff and Gunnels, John and Xanthea, Sotiris S and Ganahl, Martin and Menczer, Andor and Legeza, {\"O}rs},
  journal={arXiv preprint arXiv:2510.04795},
  year={2025},
  url={https://arxiv.org/abs/2510.04795},
  doi={10.48550/arXiv.2510.04795}
}

@article{ganahl2023density,
  title={Density matrix renormalization group with tensor processing units},
  author={Ganahl, Martin and Beall, Jackson and Hauru, Markus and Lewis, Adam GM and Wojno, Tomasz and Yoo, Jae Hyeon and Zou, Yijian and Vidal, Guifre},
  journal={PRX Quantum},
  volume={4},
  number={1},
  pages={010317},
  year={2023},
  publisher={APS},
  url={https://journals.aps.org/prxquantum/abstract/10.1103/PRXQuantum.4.010317},
  doi={10.1103/PRXQuantum.4.010317}
}

@article{fomichev2024initial,
  title={Initial state preparation for quantum chemistry on quantum computers},
  author={Fomichev, Stepan and Hejazi, Kasra and Zini, Modjtaba Shokrian and Kiser, Matthew and Fraxanet, Joana and Casares, Pablo Antonio Moreno and Delgado, Alain and Huh, Joonsuk and Voigt, Arne-Christian and Mueller, Jonathan E and others},
  journal={PRX Quantum},
  volume={5},
  number={4},
  pages={040339},
  year={2024},
  publisher={APS},
  url={https://journals.aps.org/prxquantum/abstract/10.1103/PRXQuantum.5.040339},
  doi={10.1103/PRXQuantum.5.040339}
}

@article{iaconis2024quantum,
  title={Quantum state preparation of normal distributions using matrix product states},
  author={Iaconis, Jason and Johri, Sonika and Zhu, Elton Yechao},
  journal={npj Quantum Information},
  volume={10},
  number={1},
  pages={15},
  year={2024},
  publisher={Nature Publishing Group UK London},
  url={https://www.nature.com/articles/s41534-024-00805-0},
  doi={10.1038/s41534-024-00805-0}
}

@article{kanno2025tensor,
  title={Tensor-based quantum phase difference estimation for large-scale demonstration},
  author={Kanno, Shu and Sugisaki, Kenji and Nakamura, Hajime and Yamauchi, Hiroshi and Sakuma, Rei and Kobayashi, Takao and Gao, Qi and Yamamoto, Naoki},
  journal={Proceedings of the National Academy of Sciences},
  volume={122},
  number={30},
  pages={e2425026122},
  year={2025},
  publisher={National Academy of Sciences},
  url={https://www.pnas.org/doi/abs/10.1073/pnas.2425026122},
  doi={10.1073/pnas.2425026122}
}

@article{szoldra2026scalable,
  title={Scalable Preparation of Matrix Product States with Sequential and Brick Wall Quantum Circuits},
  author={Szo{\l}dra, Tomasz and Mukherjee, Rick and Schmelcher, Peter},
  journal={arXiv preprint arXiv:2602.12042},
  year={2026},
  url={https://arxiv.org/abs/2602.12042},
  doi={10.48550/arXiv.2602.12042}
}

@article{wootters1981statistical,
  title={Statistical distance and Hilbert space},
  author={Wootters, William K},
  journal={Physical Review D},
  volume={23},
  number={2},
  pages={357},
  year={1981},
  publisher={APS},
  url={https://journals.aps.org/prd/abstract/10.1103/PhysRevD.23.357},
  doi={10.1103/PhysRevD.23.357}
}

@article{rudolph2022synergy,
  title={Synergistic pretraining of parametrized quantum circuits via tensor networks},
  author={Rudolph, Manuel S and Miller, Jacob and Motlagh, Danial and Chen, Jing and Acharya, Atithi and Perdomo-Ortiz, Alejandro},
  journal={Nature Communications},
  volume={14},
  number={1},
  pages={8367},
  year={2023},
  publisher={Nature Publishing Group UK London},
  url={https://doi.org/10.1038/s41467-023-43908-6},
doi={10.1038/s41467-023-43908-6}
}

@article{zhang2003geometric,
  title={Geometric theory of nonlocal two-qubit operations},
  author={Zhang, Jun and Vala, Jiri and Sastry, Shankar and Whaley, K Birgitta},
  journal={Physical Review A},
  volume={67},
  number={4},
  pages={042313},
  year={2003},
  publisher={APS},
  url={https://journals.aps.org/pra/abstract/10.1103/PhysRevA.67.042313},
  doi={10.1103/PhysRevA.67.042313}
}

@article{schon2005sequential,
  title={Sequential generation of entangled multiqubit states},
  author={Sch{\"o}n, Christian and Solano, Enrique and Verstraete, Frank and Cirac, J Ignacio and Wolf, Michael M},
  journal={Physical review letters},
  volume={95},
  number={11},
  pages={110503},
  year={2005},
  publisher={APS},
  url={https://journals.aps.org/prl/abstract/10.1103/PhysRevLett.95.110503},
  doi={10.1103/PhysRevLett.95.110503}
}

@article{legeza2015advanced,
  title={Advanced density matrix renormalization group method for nuclear structure calculations},
  author={Legeza, {\"O} and Veis, L and Poves, Alfredo and Dukelsky, Jorge},
  journal={Physical Review C},
  volume={92},
  number={5},
  pages={051303},
  year={2015},
  publisher={APS},
  url={https://journals.aps.org/prc/abstract/10.1103/PhysRevC.92.051303},
  doi={10.1103/PhysRevC.92.051303}
}

@article{tichai2023combining,
  title={Combining the in-medium similarity renormalization group with the density matrix renormalization group: Shell structure and information entropy},
  author={Tichai, A and Knecht, S and Kruppa, AT and Legeza, {\"O} and Moca, CP and Schwenk, A and Werner, MA and Zarand, G},
  journal={Physics Letters B},
  volume={845},
  pages={138139},
  year={2023},
  publisher={Elsevier},
  url={https://www.sciencedirect.com/science/article/pii/S0370269323004732},
  doi={10.1016/j.physletb.2023.138139}
}

@article{brokemeier2025quantum,
  title={Quantum magic and multipartite entanglement in the structure of nuclei},
  author={Br{\"o}kemeier, Florian and Hengstenberg, S Momme and Keeble, James WT and Robin, Caroline EP and Rocco, Federico and Savage, Martin J},
  journal={Physical Review C},
  volume={111},
  number={3},
  pages={034317},
  year={2025},
  publisher={APS},
  url={https://journals.aps.org/prc/abstract/10.1103/PhysRevC.111.034317},
  doi={10.1103/PhysRevC.111.034317}
}

@article{johnson2023proton,
  title={Proton-neutron entanglement in the nuclear shell model},
  author={Johnson, Calvin W and Gorton, Oliver C},
  journal={Journal of Physics G: Nuclear and Particle Physics},
  volume={50},
  number={4},
  pages={045110},
  year={2023},
  publisher={IOP Publishing},
  url={https://iopscience.iop.org/article/10.1088/1361-6471/acbece/meta},
  doi={10.1088/1361-6471/acbece}
}

@article{jaderberg2025variational,
  title={Variational preparation of normal matrix product states on quantum computers},
  author={Jaderberg, Ben and Pennington, George and Marshall, Kate V and Anderson, Lewis W and Agarwal, Abhishek and Lindoy, Lachlan P and Rungger, Ivan and Mensa, Stefano and Crain, Jason},
  journal={Physical Review Research},
  volume={8},
  number={1},
  pages={013081},
  year={2026},
  publisher={APS},
  url={https://journals.aps.org/prresearch/abstract/10.1103/m5kb-4f43},
  doi={10.1103/m5kb-4f43}
}

@article{perez2023quantum,
  title={Quantum entanglement patterns in the structure of atomic nuclei within the nuclear shell model},
  author={P{\'e}rez-Obiol, Axel and Masot-Llima, S and Romero, AM and Men{\'e}ndez, J and Rios, A and Garc{\'\i}a-S{\'a}ez, A and Juli{\'a}-D{\'\i}az, Bruno},
  journal={The European Physical Journal A},
  volume={59},
  number={10},
  pages={240},
  year={2023},
  publisher={Springer},
  url={https://link.springer.com/article/10.1140/epja/s10050-023-01151-z},
  doi={10.1140/epja/s10050-023-01151-z}
}

@article{fossez2022density,
  title={Density matrix renormalization group description of the island of inversion isotopes $^{28-33}\text{F}$},
  author={Fossez, K and Rotureau, J},
  journal={Physical Review C},
  volume={106},
  number={3},
  pages={034312},
  year={2022},
  publisher={APS},
  url={https://journals.aps.org/prc/abstract/10.1103/PhysRevC.106.034312},
  doi={10.1103/PhysRevC.106.034312}
}

@article{Vidal2003Efficient,
  title = {Efficient Classical Simulation of Slightly Entangled Quantum Computations},
  author = {Vidal, Guifr\'e},
  journal = {Phys. Rev. Lett.},
  volume = {91},
  issue = {14},
  pages = {147902},
  numpages = {4},
  year = {2003},
  month = {Oct},
  publisher = {American Physical Society},
  doi = {10.1103/PhysRevLett.91.147902},
  url = {https://link.aps.org/doi/10.1103/PhysRevLett.91.147902}
}

@article{hao2025reducing,
  title={Reducing {T} Gates with Unitary Synthesis},
  author={Hao, Tianyi and Xu, Amanda and Tannu, Swamit},
  journal={arXiv preprint arXiv:2503.15843},
  year={2025},
  url={https://arxiv.org/abs/2503.15843},
  doi={10.48550/arXiv.2503.15843}
}

@article{ge2019faster,
  title={Faster ground state preparation and high-precision ground energy estimation with fewer qubits},
  author={Ge, Yimin and Tura, Jordi and Cirac, J Ignacio},
  journal={Journal of Mathematical Physics},
  volume={60},
  number={2},
  year={2019},
  publisher={AIP Publishing},
  url={https://pubs.aip.org/aip/jmp/article-abstract/60/2/022202/918222/Faster-ground-state-preparation-and-high-precision?redirectedFrom=fulltext},
  doi={10.1063/1.5027484}
}

@article{orus2019tensor,
  title={Tensor networks for complex quantum systems},
  author={Or{\'u}s, Rom{\'a}n},
  journal={Nature Reviews Physics},
  volume={1},
  number={9},
  pages={538--550},
  year={2019},
  publisher={Nature Publishing Group UK London},
  url={https://www.nature.com/articles/s42254-019-0086-7},
  doi={10.1038/s42254-019-0086-7}
}

@article{gibbs2025learning,
  title={Learning Circuits with Infinite Tensor Networks},
  author={Gibbs, Joe and Cincio, Lukasz},
  journal={arXiv preprint arXiv:2506.02105},
  year={2025},
url={https://arxiv.org/abs/2506.02105},
doi={10.48550/arXiv.2506.02105}
}

@article{gibbs2024deep,
  title={Deep Circuit Compression for Quantum Dynamics via Tensor Networks},
  author={Gibbs, Joe and Cincio, Lukasz},
  journal={Quantum},
  volume={9},
  pages={1789},
  year={2025},
  publisher={Verein zur F{\"o}rderung des Open Access Publizierens in den Quantenwissenschaften},
url={https://quantum-journal.org/papers/q-2025-07-09-1789/},
doi={10.22331/q-2025-07-09-1789}
}

@article{lee2023evaluating,
  title={Evaluating the evidence for exponential quantum advantage in ground-state quantum chemistry},
  author={Lee, Seunghoon and Lee, Joonho and Zhai, Huanchen and Tong, Yu and Dalzell, Alexander M and Kumar, Ashutosh and Helms, Phillip and Gray, Johnnie and Cui, Zhi-Hao and Liu, Wenyuan and others},
  journal={Nature Communications},
  volume={14},
  number={1},
  pages={1952},
  year={2023},
  publisher={Nature Publishing Group UK London},
  url={https://www.nature.com/articles/s41467-023-37587-6},
  doi={10.1038/s41467-023-37587-6}
}

@article{chai2025resource,
  title={Resource-Efficient Simulations of Particle Scattering on a Digital Quantum Computer},
  author={Chai, Yahui and Gibbs, Joe and Pascuzzi, Vincent R and Holmes, Zo{\"e} and K{\"u}hn, Stefan and Tacchino, Francesco and Tavernelli, Ivano},
  journal={arXiv preprint arXiv:2507.17832},
  year={2025},
  url={https://arxiv.org/abs/2507.17832},
  doi={10.48550/arXiv.2507.17832}
}

@article{kitaev1995quantum,
  title={Quantum measurements and the Abelian stabilizer problem},
  author={Kitaev, A Yu},
  journal={arXiv preprint quant-ph/9511026},
  year={1995},
  url={https://arxiv.org/abs/quant-ph/9511026},
  doi={10.48550/arXiv.quant-ph/9511026}
}

@article{bhoy2024shell,
  title={Shell-model study of $^{58}\text{Ni}$ using quantum computing algorithm},
  author={Bhoy, Bharti and Stevenson, Paul},
  journal={New Journal of Physics},
  volume={26},
  number={7},
  pages={075001},
  year={2024},
  publisher={IOP Publishing},
  url={https://iopscience.iop.org/article/10.1088/1367-2630/ad5756/meta},
  doi={10.1088/1367-2630/ad5756}
}

@article{lin2021real,
  title={Real-and imaginary-time evolution with compressed quantum circuits},
  author={Lin, Sheng-Hsuan and Dilip, Rohit and Green, Andrew G and Smith, Adam and Pollmann, Frank},
  journal={PRX Quantum},
  volume={2},
  number={1},
  pages={010342},
  year={2021},
  publisher={APS},
  doi = {10.1103/PRXQuantum.2.010342},
  url = {https://doi.org/10.1103/PRXQuantum.2.010342}
}

@article{fishman2022itensor,
  doi={10.21468/SciPostPhysLectNotes.54},
  url={https://scipost.org/10.21468/SciPostPhysLectNotes.54},
  title={The ITensor software library for tensor network calculations},
  author={Fishman, Matthew and White, Steven and Stoudenmire, Edwin},
  journal={SciPost Physics Codebases},
  pages={004},
  year={2022}
}

@article{lin2025bounds,
  title={Bounds on a Wavefunction Overlap with Hamiltonian Eigen-states: Performance Guarantees for the Quantum Phase Estimation Algorithm},
  author={Lin, Junan and Izmaylov, Artur F},
  journal={arXiv preprint arXiv:2503.12224},
  year={2025},
  url={https://arxiv.org/abs/2503.12224},
  doi={10.48550/arXiv.2503.12224}
}

@article{ross2014optimal,
  author = {Ross, Neil J. and Selinger, Peter},
  title = {Optimal ancilla-free Clifford+{T} approximation of z-rotations},
  year = {2016},
  issue_date = {September 2016},
  publisher = {Rinton Press, Incorporated},
  address = {Paramus, NJ},
  volume = {16},
  number = {11–12},
  issn = {1533-7146},
  journal = {Quantum Info. Comput.},
  month = sep,
  pages = {901–953},
  numpages = {53},
  doi={0.26421/qic16.11-12-1},
  url={ https://doi.org/10.26421/qic16.11-12-1 }
}

@article{miller2025simulation,
  title={Simulation of Fermionic circuits using Majorana Propagation},
  author={Miller, Aaron and Favre, Joachim and Holmes, Zo{\"e} and Salehi, {\"O}zlem and Chakraborty, Rahul and Nyk{\"a}nen, Anton and Zimboras, Zoltan and Glos, Adam and Garc{\'\i}a-P{\'e}rez, Guillermo},
  journal={arXiv preprint arXiv:2503.18939},
  year={2025},
  url={https://doi.org/10.48550/arXiv.2503.18939}, 
doi={10.48550/arXiv.2503.18939}
}

@article{bosyk2014geometric,
  title={Geometric formulation of the uncertainty principle},
  author={Bosyk, Gustavo Mart{\'\i}n and Os{\'a}n, Trist{\'a}n Mart{\'\i}n and Lamberti, Pedro Walter and Portesi, Mariela},
  journal={Physical Review A},
  volume={89},
  number={3},
  pages={034101},
  year={2014},
  publisher={APS},
  url={https://journals.aps.org/pra/abstract/10.1103/PhysRevA.89.034101},
  doi={10.1103/PhysRevA.89.034101}
}

@article{ran2020encoding,
  title={Encoding of matrix product states into quantum circuits of one-and two-qubit gates},
  author={Ran, Shi-Ju},
  journal={Physical Review A},
  volume={101},
  number={3},
  pages={032310},
  year={2020},
  publisher={APS},
  url={https://doi.org/10.1103/PhysRevA.101.032310}
}

@article{le2025riemannian,
  title={Riemannian quantum circuit optimization based on matrix product operators},
  author={Le, Isabel Nha Minh and Sun, Shuo and Mendl, Christian B},
  journal={Quantum},
  volume={9},
  pages={1833},
  year={2025},
  publisher={Verein zur F{\"o}rderung des Open Access Publizierens in den Quantenwissenschaften},
  url = {https://quantum-journal.org/papers/q-2025-08-27-1833/},
  doi = {10.22331/q-2025-08-27-1833}
}

@incollection{chan2008introduction,
  title={An introduction to the density matrix renormalization group ansatz in quantum chemistry},
  author={Chan, Garnet Kin-Lic and Dorando, Jonathan J and Ghosh, Debashree and Hachmann, Johannes and Neuscamman, Eric and Wang, Haitao and Yanai, Takeshi},
  booktitle={Frontiers in quantum systems in chemistry and physics},
  pages={49--65},
  year={2008},
  publisher={Springer},
  url={https://link.springer.com/chapter/10.1007/978-1-4020-8707-3_4},
  doi={10.1007/978-1-4020-8707-3_4}
}

@article{jamet2023anderson,
  title={Anderson impurity solver integrating tensor network methods with quantum computing},
  author={Jamet, Fran{\c{c}}ois and Lindoy, Lachlan P and Rath, Yannic and Lenihan, Connor and Agarwal, Abhishek and Fontana, Enrico and Simkovic, Fedor and Martin, Baptiste Anselme and Rungger, Ivan},
  journal={APL Quantum},
  volume={2},
  number={1},
  year={2025},
  publisher={AIP Publishing},
  url={https://pubs.aip.org/aip/apq/article/2/1/016121/3336642/Anderson-impurity-solver-integrating-tensor},
  doi={10.1063/5.0245488}
}

@article{causer2023scalable,
  title={Scalable simulation of nonequilibrium quantum dynamics via classically optimized unitary circuits},
  author={Causer, Luke and Jung, Felix and Mitra, Asimpunya and Pollmann, Frank and Gammon-Smith, Adam},
  journal={Physical Review Research},
  volume={6},
  number={3},
  pages={033062},
  year={2024},
  publisher={APS},
  url={https://doi.org/10.1103/PhysRevResearch.6.033062},
  doi={10.1103/PhysRevResearch.6.033062}
}

@article{shirakawa2021automatic,
  title={Automatic quantum circuit encoding of a given arbitrary quantum state},
  author={Shirakawa, Tomonori and Ueda, Hiroshi and Yunoki, Seiji},
  journal={Physical Review Research},
  volume={6},
  number={4},
  pages={043008},
  year={2024},
  publisher={APS},
  url={https://journals.aps.org/prresearch/abstract/10.1103/PhysRevResearch.6.043008},
  doi={10.1103/PhysRevResearch.6.043008}
}

@article{evenbly2009algorithms,
  title={Algorithms for entanglement renormalization},
  author={Evenbly, Glen and Vidal, Guifr{\'e}},
  journal={Physical Review B—Condensed Matter and Materials Physics},
  volume={79},
  number={14},
  pages={144108},
  year={2009},
  publisher={APS},
  url={https://doi.org/10.1103/PhysRevB.79.144108},
  doi={10.1103/PhysRevB.79.144108}
}

@article{anselme2024combining,
  title={Combining matrix product states and noisy quantum computers for quantum simulation},
  author={Anselme Martin, Baptiste and Ayral, Thomas and Jamet, Fran{\c{c}}ois and Ran{\v{c}}i{\'c}, Marko J and Simon, Pascal},
  journal={Physical Review A},
  volume={109},
  number={6},
  pages={062437},
  year={2024},
  publisher={APS},
  url={https://doi.org/10.1103/PhysRevA.109.062437},
  doi={10.1103/PhysRevA.109.062437}
}

@article{lin2025utility,
  title={Utility-Scale Quantum State Preparation: Classical Training using Pauli Path Simulation},
  author={Lin, Cheng-Ju and Gharibyan, Hrant and Su, Vincent P},
  journal={arXiv preprint arXiv:2510.02428},
  year={2025},
  url={https://arxiv.org/abs/2510.02428},
  doi={10.48550/arXiv.2510.02428}
}

@article{rudolph2025pauli,
  title={Pauli Propagation: A Computational Framework for Simulating Quantum Systems},
  author={Rudolph, Manuel S and Jones, Tyson and Teng, Yanting and Angrisani, Armando and Holmes, Zo{\"e}},
  journal={arXiv preprint arXiv:2505.21606},
  year={2025},
  url={https://arxiv.org/abs/2505.21606},
doi={10.48550/arXiv.2505.21606}
}
